\documentclass[nohyper]{JHEP3}
\usepackage{amsmath,amssymb,amsfonts}
\usepackage{slashed}
\usepackage{graphicx}

\newcommand{\ket}[1]{|#1\rangle}

\newcommand{\bracket}[2]{\langle #1|#2\rangle}
\newcommand{\matelem}[3]{\langle #1|#2|#3\rangle}
\newcommand{\measure}[1]{\mathcal{D}#1}

\newcommand{\figI}{
\,\raisebox{-0.1cm}{\includegraphics[height=.4cm]{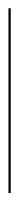}}\,
}

\newcommand{\figII}{
\,\raisebox{-0.1cm}{\includegraphics[height=.4cm]{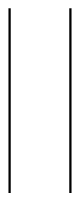}}\,
}

\newcommand{\figIX}{
\,\raisebox{-0.1cm}{\includegraphics[height=.4cm]{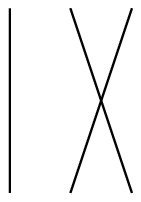}}\,
}

\newcommand{\figLtwo}{
\,\raisebox{-0.1cm}{\includegraphics[height=.4cm]{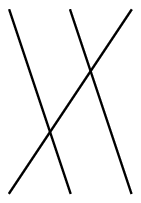}}\,}

\newcommand{\figX}{
\,\raisebox{-0.1cm}{\includegraphics[height=.4cm]{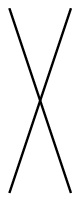}}\,
}

\preprint{
 ITP-UU-08/70 \hfill\\
 NIKHEF-2008-33\hfill\\
 ITFA-2008-44\hfill\\
 }

\title{Path integral approach to eikonal and next-to-eikonal exponentiation}

\author{Eric Laenen\\
  ITFA, University of Amsterdam,
  Valckenierstraat 65, 1018 XE Amsterdam, \\
  ITF, Utrecht University, Leuvenlaan 4, 3584 CE Utrecht\\
  Nikhef Theory Group,  Kruislaan 409, 1098 SJ Amsterdam, The Netherlands\\
  E-mail: \email{Eric.Laenen@nikhef.nl}}
\author{Gerben Stavenga\\
  Institute for Theoretical Physics, Utrecht University, Leuvenlaan 4, 3584 CE Utrecht, The Netherlands\\
 E-mail: \email{stavenga@phys.uu.nl}}
\author{Chris D. White\\
  Nikhef Theory Group,  Kruislaan 409, 1098 SJ Amsterdam, The Netherlands\\
  E-mail: \email{cwhite@nikhef.nl}}

\abstract{We approach the issue of exponentiation of soft gauge boson
  corrections to scattering amplitudes from a path integral point of
  view. We show that if one represents the amplitude as a first
  quantized path integral in a mixed coordinate-momentum space
  representation, a charged particle interacting with a soft gauge
  field is represented as a Wilson line for a
  semi-infinite line segment, together with calculable
  fluctuations. Combining such line segments, we show that exponentiation
  in an abelian field theory follows immediately from standard path-integral
  combinatorics. In the non-abelian case, we consider color singlet hard interactions
  with two outgoing external lines, and obtain a new viewpoint for
  exponentiation in terms of ``webs'', with a closed form solution for their corresponding
  color factors. We investigate and clarify the structure of next-to-eikonal corrections.
}

\begin{document}
\section{Introduction}
Higher order corrections arising from soft gauge bosons in perturbative gauge theory,
 be they real or virtual, have been the subject of many
investigations. Such radiation
generically leads to series of perturbative contributions to differential cross-sections of 
the form $\alpha^n \log^m(\xi)/\xi$, where $\alpha$ is the coupling constant of
the gauge theory, and $\xi$ is related to the energy carried away by 
the soft particles. In the soft limit
$\xi\rightarrow 0$ (in which the eikonal approximation may be taken), 
it becomes necessary to resum
these terms to all orders in perturbation theory, 
as has been achieved by a variety of methods
 \cite{Sterman:1986aj,Catani:1989ne,Catani:1984dp,Korchemsky:1993uz,Korchemsky:1994jb,Contopanagos:1997nh,Forte:2002ni,Becher:2006nr}. 
Central to resummation is the exponentiation of eikonalized soft gauge boson
corrections and it has been shown
for both abelian and non-abelian gauge theory that this indeed occurs
\cite{Yennie:1961ad,Gatheral:1983cz,Frenkel:1984pz,Sterman:1981jc}.

\begin{figure}
\begin{center}
\scalebox{0.6}{\includegraphics{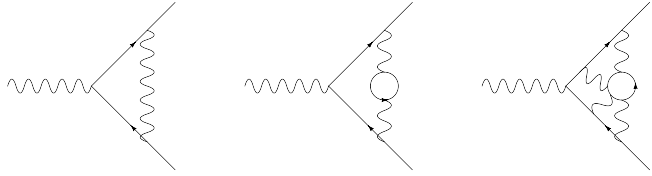}}
\caption{Examples of connected diagrams $G_c$ of soft emissions
between hard outgoing particle legs in abelian perturbation theory.}
\label{exp1}
\end{center}
\end{figure}

\begin{figure}
\begin{center}
\scalebox{0.6}{\includegraphics{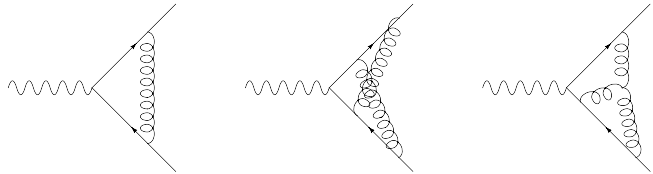}}
\caption{Examples of webs of soft emissions between external lines in 
non-abelian perturbation theory.}
\label{exp2}
\end{center}
\end{figure}

To form an exponential series for a cross-section, both the matrix element and the phase space 
must exhibit an appropriate factorized structure. 
The statement for the matrix elements rests upon a
thorough analysis of the general structure of higher order
diagrams. The result for an amplitude ${\cal A}$ in abelian 
gauge theory is simply expressed in the eikonal approximation as
\begin{equation}
{\cal A}={\cal A}_0\exp\left[\sum G_c\right],
\label{Aexp}
\end{equation}
where ${\cal A}$ is the amplitude without soft radiation containing a number
of hard outgoing {\it external lines}, and 
the sum in the exponent is over connected diagrams $G_c$
between the external lines, with the Born contribution 
${\cal A}_0$ factored out. Examples are shown for the case
of hard production of a particle-antiparticle pair in figure 
\ref{exp1}.
The non-abelian case is complicated by the nontrivial color structure 
of the Feynman diagrams at successive orders in perturbation theory.
Nevertheless, in the case of two external lines\footnote{In the case of more than two
external lines, the structure is more complicated.}, exponentiation still holds provided one generalizes 
Eq.~(\ref{Aexp}) to
\begin{equation}
{\cal A}={\cal A}_0\exp\left[\sum \bar{C}_{W} W\right].
\label{Aexp2}
\end{equation}
Here $W$ are so-called {\it webs}, and are diagrams which are 
two-eikonal irreducible. That is, one cannot partition a web into webs
of lower order by cutting both external lines exactly 
once~\cite{Gatheral:1983cz,Frenkel:1984pz,Sterman:1981jc}\footnote{
See~\cite{Berger:2003zh} for a pedagogical exposition.}. Examples 
are shown in figure \ref{exp2}. 
Each web has an associated color factor $\bar{C}_W$ 
which is not the same as the normal color factor $C_W$ associated 
with the web graph. The color factors $\bar{C}_W$ are given 
in terms of $C_W$ by an iterative relation to all orders 
in perturbation theory.

The nature of Eq.(~\ref{Aexp}) in terms of disconnected diagrams
is reminiscent of another well-known property of quantum field 
theory, namely the exponentiation of disconnected Feynman diagrams
in terms of connected ones. 
This latter property most naturally emerges using path
integral methods (see appendix \ref{expproof} for a brief proof), 
and thus the suggestion arises of whether it is
possible to relate the exponentiation of soft radiative corrections
to the gauge theory path integral. The aim of this paper is to show that 
this is indeed the case, and the result is important for a number of reasons.
Firstly, it provides a new perspective on exponentiation. Secondly, 
it allows one to straightforwardly explore which
properties of exponentiation survive at next-to-eikonal limit and 
beyond (i.e. corresponding to subleading terms in $\xi$ above)\footnote{We 
only consider matrix elements in this paper. Exponentiation in differential 
cross-sections also depends upon factorization of the 
multiple particle phase space, and is deferred for further study.}.
Although at next-to-eikonal order fermion emissions also contribute (leading to flavor-changing effects),
we restrict ourselves here to gluon emissions. 

The essential idea of our approach is as follows. One first relates the 
field theory path integral for a particle interacting with 
a gauge field to a first-quantized path integral with respect to the 
particle. That is, the external lines become worldlines of particles
in quantum mechanics (rather than quantum field theory). Here we utilize the
techniques of \cite{Strassler:1992zr,Schmidt:1994zj,vanHolten:1995ds} which
have originally been applied in a different context (that of constructing string
theoretical analogues of fixed order field theory amplitudes
~\cite{Bern:1991an,Bern:1991aq}). A first-quantized approach to Sudakov resummation has
also appeared in \cite{Karanikas:2002sy}. We will see explicitly 
that in this representation the eikonal limit corresponds to the radiating 
particles moving classically, and next-to-eikonal terms originate from fluctuations 
around the classical path. The soft radiation emission vertices can then 
be interpreted as interactions of the gauge field with a source, such that 
individual emission vertices form disconnected diagrams. Then exponentiation
of eikonal corrections follows naturally from usual combinatoric properties
of the path integral. 

In the non-abelian case, exponentiation is complicated by the fact that
vertices for the emissions of gluons do not commute. However, one can rephrase
the problem using the {\it replica trick} of statistical physics (see e.g.~\cite{Replica}),
such that a subset of diagrams arises which exponentiate. These are then precisely the webs 
of~\cite{Gatheral:1983cz,Frenkel:1984pz}. Furthermore, we provide an explicit closed form solution
for the modified color factors, given in terms of normal (rather than modified) color weights.

Our formalism allows one to straightforwardly consider subleading effects with respect to the
eikonal limit, and we classify the possible next-to-eikonal corrections. This can be divided into
a subset which exponentiate (involving NE generalizations of the webs discussed above), and a set
of remainder terms which do not formally exponentiate, but have an iterative structure in that
each order of the perturbation expansion is sufficient to generate the next order.

Earlier attempts to include certain sub-eikonal effects 
were done in practical implementations of Sudakov resummation, 
mostly in view of gauging the theoretical uncertainty of the
resummation~\cite{Kramer:1996iq}.  Typically, this involved including
subleading terms in the collinear evolution kernel in the
resummation, which is particularly appealing for Drell-Yan, Higgs production and related
cross sections \cite{Kramer:1996iq,Catani:2001ic,Harlander:2001is,Catani:2003zt,Kidonakis:2007ww,Basu:2007nu}. 
More recently a study was performed \cite{Laenen:2008ux} based on a proposal in 
Ref.~\cite{Dokshitzer:2005bf}.

In the rest of this introduction we review
the derivation of the path-integral representation of propagators in quantum
field theory, for both scalar and spinor particles and in the presence and absence
of an abelian gauge field. These results will be used 
in section 2 to demonstrate the exponentiation of soft radiative corrections
in the presence of an abelian gauge field. We also consider generalization of the
results to beyond the eikonal limit, and classify the resulting corrections into a subset
which exponentiate and a remainder term which mixes with these at next-to-eikonal order.
In section 3 we consider the extension to non-abelian gauge fields, recovering 
the properties of webs and again examining corrections to the eikonal limit. We 
conclude in section 4, and some technical details are presented in the appendices.
\subsection{Propagators as first quantized path integrals}
\begin{figure}
\begin{center}
\scalebox{0.9}{\includegraphics{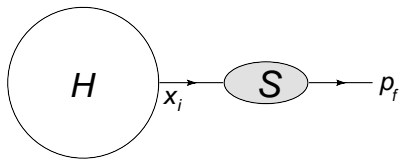}}
\caption{Anatomy of an external line, considered throughout this paper as 
emerging from a hard interaction at 4-position $x_i$ at time $t=0$, and having
momentum $p_f$ at final time $T$. The propagator for the external line contains 
the effects of soft radiation.}
\label{eikline}
\end{center}
\end{figure}

Throughout this paper we will be concerned with {\it external lines}, namely hard
external particles susceptible to soft radiation emission. A given external line
is created from a hard interaction process at time $t=0$ at space-time point $x_i$, 
and and has a momentum $p_f$ at some final time $T\rightarrow\infty$ (see 
figure~\ref{eikline}). 
Soft radiation
corrections enter the two-point function for the emitting particle i.e. the 
propagator for the external line. In the following subsections, we review the 
representation of field theory propagators in terms of first quantized 
path integrals~\cite{Strassler:1992zr,Schmidt:1994zj,vanHolten:1995ds},
which will later on be used in the derivation of matrix element exponentiation
at eikonal order and beyond. We begin with the simplest case, that of a free scalar particle.
\subsubsection{Free scalar particle}
\label{sec:scalar-particle}
The propagator for a free scalar particle between 4-positions $x$ and $y$ is the Green's
function for the Klein-Gordon equation
\begin{equation}
  \label{eq:3}
  i(S-i\varepsilon) \Delta_F(x,y)=\delta^{(d)}(x,y), \qquad S = (-\Box_x+m^2),
\end{equation}
where $S, \Delta_F$ are Hermitian operators working on the Hilbert
space $\mathcal{H}$ of square integrable functions of space-time, and we adopt the standard
Feynman $i\varepsilon$ prescription. Note that we are using the metric $(-,+,+,+)$. 
Schematically the propagator may be written as
\begin{equation}
  \label{eq:4}
  \Delta_F = [i\left(S-i\varepsilon\right)]^{-1},
\end{equation}
where the inverse operator can be defined via an inverse Fourier transform
from momentum space. The usual representation then reads
\begin{equation}
  \label{eq:1}
  \Delta_F(x,y) = -i\int \frac{d^dp}{(2\pi)^d} \frac{e^{i p\cdot (y-x)}}{p^2+m^2-i\varepsilon}\,.
\end{equation}
To derive this expression using a first-quantized path integral, let
us first write the inverse Klein-Gordon operator using the Schwinger
representation
\begin{equation}
  \label{eq:2}
  -i (S-i\varepsilon)^{-1}=\frac12 \int_0^\infty dT e^{-i \frac12(S-i\varepsilon) T}
\end{equation}
where the $i\varepsilon$ now ensures the convergence of the integral.
The integrand contains the exponential $U(T)=e^{-i \frac12 S T}$, 
which is a unitary operator acting on the Hilbert space ${\cal H}$ and
satisfying the Schr\"odinger equation
\begin{equation}
  \label{eq:5}
  i\frac{d}{dT} U(T)= \hat{H} U(T), \quad U(0)=I, \quad \hat{H}=\frac12 S.
\end{equation}
We can therefore interpret $\hat{H}$ as the Hamiltonian operator of
a quantum system, with internal time coordinate $T$. Given we are considering
external lines as shown in figure~\ref{eikline}, we must calculate the
expectation value of the evolution operator $U$ between a state of 
definite position (at time $t=0$) and definite momentum (at time $t=T$).
To do this we introduce states $\ket{x}$ and $\ket{p}$ in the Hilbert
space $\mathcal{H}$, and write the Hamiltonian as
\begin{equation}
\hat{H}(\hat{x},\hat{p})=\sum_{n=0}^\infty \hat{p}_{\mu_1}\ldots
\hat{p}_{\mu_n} \,H^{\mu_1\ldots\mu_n}_{\nu_1\ldots\nu_n} \,\hat{x}^{\nu_1\ldots\nu_n},
\label{Hexp}
\end{equation}
where $\hat{x}$ and $\hat{p}$ are the position and momentum operators whose
continuous eigenstates are $\ket{x}$ and $\ket{p}$ respectively. Note we
have expressed $\hat{H}$ in Weyl ordered form, with all momentum operators
to the left of position operators. One then finds, for small time separations
$\Delta t$
\begin{equation}
  \label{eq:7}
  \matelem{p}{e^{-i H \Delta t}}{x}=e^{-i H(p,x) \Delta t + O[(\Delta t)^2]}\bracket{p}{x},
\end{equation}
where $H(p,x)$ is the c-number obtained by replacing the operators on the right-hand-side
of Eq.~(\ref{Hexp}) with their corresponding variables. Slicing the time variable into
$N$ steps of duration $\Delta t$ and inserting a complete set of both position
and momentum states at each step, Eq.~(\ref{eq:7}) becomes
\begin{equation}
\int dx_1\ldots dx_{N}\int dp_0\ldots dp_{N-1}\exp\left[-i\sum_{k=0}^{N-1} H(p_k,x_{k})
\Delta t\right]\prod_{k=0}^N \bracket{p_k}{x_k}\prod_{k=0}^{N-1}\bracket{x_{k+1}}{p_k}.
\label{path1}
\end{equation}
Using the normalization of the basis states
\begin{equation}
\bracket{x}{p}=\frac{e^{i x p}}{(2\pi)^d},
\label{norm}
\end{equation}
where $d$ is the number of space-time dimensions, the continuum limit of 
Eq.(~\ref{path1}) is
\begin{equation}
  \label{eq:8}
  \matelem{p_f}{U(T)}{x_i}=\int_{x(0)=x_i}^{p(T)=p_f} \measure{p} \measure{x} \,
  \exp\left[-i p(T) x(T) + i\int_0^T dt (p\dot{x}-H(p,x)) \right].
\end{equation}
We have absorbed factors of $2\pi$ into the measure, and made the boundary conditions
explicit. This is the well-known path-integral result for the evolution operator
sandwiched between initial and final position states, with an additional term in the 
exponent involving $p(T)x(T)$ arising from considering a final state of given momentum,
rather than position.

For the present case of a free massive scalar, the Hamiltonian function is given by
\begin{equation}
  \label{eq:9}
  H(p)  = \frac{1}{2}\left(p^2+m^2\right).
\end{equation}
We can perform the path integrations over $p(t)$ and $x(t)$ by expanding 
around the classical solution of the equations of motion, given by
\begin{equation}
  \label{eq:6}
  p(t)=p_f+p'(t), \quad x(t)=x_i+p_f t+x'(t).
\end{equation}
The boundary conditions imply $p'(T)=0$ and $x'(0)=0$, and without confusion
we can drop the primed notation from now on. Substituting Eq.~(\ref{eq:6})
into Eq.~(\ref{eq:8}) gives
\begin{equation}
  \label{eq:10}
  \matelem{p_f}{U(T)}{x_i}=e^{-ip_f x_i-i\frac12(p_f^2+m^2)T}\int_{x(0)=0}^{p(T)=0} \measure{p} \measure{x} \,
  e^{i\int_0^T dt (p\dot{x}-\frac12 p^2) }.
\end{equation}
One can now perform the path integral as the continuum limit of a product of 
Gaussian integrals in the intermediate position and momentum variables. The
measure is such that this gives unity, and one therefore finds
\begin{equation}
  \label{eq:11}
  \matelem{p_f}{U(T)}{x_i}=e^{-ip_f x_i-\frac12 i(p_f^2+m^2)T}.
\end{equation}
The momentum space propagator $\tilde{\Delta}_F$ is found by substituting Eq.~(\ref{eq:11})
into Eq.~(\ref{eq:2}), and one finds
\begin{equation}
\tilde{\Delta}_F(p_f^2)=\frac12\int_0^\infty dT\frac{\matelem{p_f}{U(T)}{x_i}}{\bracket{p_f}{x_i}}=-\frac{i}
{p_f^2+m^2-i\varepsilon},
\label{propmom}
\end{equation}
in agreement with Eq.~(\ref{eq:1}). Having reviewed the relationship between propagators
and path integrals in a simple case, we now consider the extension to a scalar particle
interacting with a gauge field.
\subsection{Scalar particle in an abelian background gauge field}
We consider a charged scalar particle in an abelian
background gauge field. Such a system is described by the
generating functional
\begin{equation}
  \label{eq:15}
  Z[J^*, J]=\int \measure{\phi^*}\measure{\phi}\, \exp\Big[i\int d^dx 
  \left(\phi^*(D_\mu D^\mu-m^2+i\varepsilon)\phi+J^*\phi+\phi^*J\right)\Big]\,,
\end{equation}
where $J,J^*$ are sources for the complex scalar field, and
$D_\mu=\partial_\mu-iA_\mu$.  By completing the square and defining
$S=(-D_\mu D^\mu+m^2)$ we can write this as
\begin{equation}
  Z[J^*, J]=\int \measure{\phi^*}\measure{\phi}\,
  \exp\left[i\int d^dx \left(-\phi^*(S-i\varepsilon)\phi-J^*(S-i\varepsilon)^{-1}J\right)\right]\,.
\end{equation}
The propagator is given by the inverse of operator quadratic in $\phi$, $\phi^*$,
which gives Eq.~(\ref{eq:4}) as before. Using the fact that $p_\mu=-i\partial_\mu$, we can
write the operator $S$ in normal form (i.e. with all momenta operators on the left-hand-side) 
as
\begin{equation}
  S=(p-A)^2+m^2=p^2-p\cdot A-A\cdot p+A^2+m^2=p^2-2 p\cdot A-i(\partial\cdot A)+A^2+m^2.
\label{SA}
\end{equation}
Now defining the Hamiltonian operator $H=\frac{1}{2}S$ as before, one may carry out the
manipulations of the previous section to obtain the first-quantized path integral 
representation of the evolution operator sandwiched between the external line position and
momentum states
\begin{multline}
  \label{eq:14}
  \matelem{p_f}{U(T)}{x_i}=\int_{x(0)=x_i}^{p(T)=p_f} \measure{p}
  \measure{x}
  \exp\Big[-i p(T) x(T) + i\int_0^T dt (p\dot{x}-\frac12 (p^2+m^2)+p\cdot A\\+\frac i2\partial\cdot A-\frac12 A^2) \Big]\,. \\
\end{multline}
This differs from the free particle case due to the presence of the
gauge field in the exponent. When the strength of the gauge field is 
weak, the classical path of the emitting particle is well approximated
by the free particle solution of Eq.~(\ref{eq:6})\footnote{We will formalize
this statement in section \ref{sec:next-eikon-expon} when we discuss next-to-eikonal exponentiation.}.
One then finds
\begin{equation}
  \label{eq:13}
  \matelem{p_f}{U(T)}{x_i}=e^{-ip_f x_i-i\frac12 (p_f^2+m^2)T} f(T),
\end{equation}
where
\begin{multline}
\label{f}
  f(T)=\int_{x(0)=0}^{p(T)=0}\measure{p} \measure{x} \, \exp\Big[i\int_0^T
  dt (p\dot{x}-\frac12 p^2+ (p_f+p)\cdot A(x_i+p_f t+x)\\+ \frac
  i2\partial\cdot A(x_i+p_f t+x)-A^2(x_i+p_f t+x)) \Big].\\
\end{multline}
Again the boundary conditions have again been made explicit, and we have dropped
the primes on the quantities defined in Eq.~(\ref{eq:6}). 
\subsection{Spinor particle}
We now consider the case of an emitting fermion in the presence of a background gauge field.
The system is described by the generating functional
\begin{equation}
Z[\eta,\bar{\eta}]=\int\measure{\bar{\psi}}\measure{\psi}\exp\left[i\int d^dx\left(\bar{\psi}
\,(\slashed{D}-m)\psi+\bar{\eta}\psi+\bar{\psi}{\eta}\right)\right],
\label{Zferm}
\end{equation}
where $\bar{\eta}$, $\eta$ are Grassmann-valued source fields\footnote{Recall we use the metric $(-,+,+,+)$ throughout.}. 
The momentum space propagator is then given by
\begin{equation}
\Delta_F=\frac{1}{-i({\slashed D}-m)}=({\slashed D}+m)\frac{1}{i[(-i{\slashed D})^2+m^2]}.
\label{fermprop}
\end{equation}
Now we define $S=(-i{\slashed D})^2+m^2$, by analogy with the scalar case.
Using the standard trick
\begin{equation}
\gamma^\mu\gamma^\nu=\frac12 \{\gamma^\mu,\gamma^\nu\}+\frac12 [\gamma^\mu,\gamma^\nu],
\label{gamtrick}
\end{equation}
we can rewrite 
\begin{equation}
(-i{\slashed D})^2+m^2=p^2-2p\cdot A-i\partial\cdot A +A^2+m^2-\sigma^{\mu\nu}F_{\mu\nu},
\label{SAferm3}
\end{equation}
where $\sigma^{\mu\nu}=-\frac{i}{4}[\gamma^{\mu},\gamma^{\nu}]$ are the generators of the Lorentz group,
and $F_{\mu\nu}$ is the field strength tensor for the gauge field.
Carrying out the path integral manipulations as in the scalar case yields
\begin{equation}
  \label{fermpath}
  \matelem{p_f}{U(T)}{x_i}=\int_{x(0)=x_i}^{p(T)=p_f} \measure{p} 
  \measure{x} P e^{-i p(T) x(T) + i\int_0^T dt (p\dot{x}-\frac12 
    (p^2+m^2)+p\cdot A+\frac i2\partial\cdot A-\frac12 A^2)+\frac12 
    \sigma^{\mu\nu} F_{\mu\nu} },
\end{equation}
The representation (\ref{fermpath}) is almost identical to the case of a scalar eikonal
line, apart from the coupling to the field strength tensor. This is not surprising,
as it is well known~\cite{Strassler:1992zr,Brink:1976sc} that one can
cast fermion actions into a second-order form that gives rise to scalar-like 
vertices supplemented by additional seagull vertices involving couplings to the
field strength (see e.g.~\cite{Morgan:1995te,Lam:1993cu}). The latter correspond
physically to the magnetic moment of the spinning emitting particle, and thus
have no analogue in the scalar case.

In this introduction, we have reviewed the representation of particle propagators, including  
the possible presence of an abelian background gauge field, as first-quantized path integrals. 
The phase space variables $x$ and $p$ correspond to the position and momentum of the 
emitting particle, and the classical path is interpreted as an eikonal line. In the following
section we formalize these statements, and show how the above representations can be used to 
derive the exponentiation of soft radiative corrections.

\section{Soft emissions in scattering processes}
\label{sec:soft-emiss-scatt}
\begin{figure}
\begin{center}
\scalebox{0.8}{\includegraphics{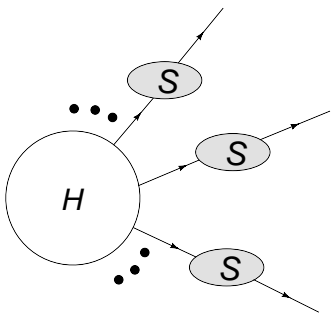}}
\caption{The factorized form of the Green's functions of Eq.~(\ref{eq:16}), where 
$H$ is a hard interaction with $n$ outgoing external lines, and $S$ is a propagator for the eikonal particle 
in the presence of the background soft gauge field which, after the path integral over $A^\mu_s$, generates connections
between the external lines.}
\label{Gdef}
\end{center}
\end{figure}

We now turn to the description of soft radiation from external lines, considering Green's functions
having the form shown schematically in figure \ref{Gdef}, and consisting of a {\it hard interaction}
$H(x_1,\ldots,x_n)$ with external lines emerging at positions $\{x_i\}$. This is a sum of subdiagrams 
containing gauge boson modes of as yet unspecified momentum. Each external line has a propagator associated with 
it summing the effect of soft gauge boson emission, and we call such diagrams {\it eikonally factorized}.

This is based upon the general analysis of \cite{Sterman:1987aj}, which characterizes the regions of 
infrared sensitivity in Feynman diagrams in a number of scattering processes. The proof that 
soft radiation contributions exponentiate now amounts to showing two things.
Firstly, that for the eikonally factorized diagrams defined above (and shown in Fig.~\ref{Gdef}), the 
contributions from soft radiation on the outgoing external lines exponentiates. Secondly, that all leading soft
radiation terms originate from diagrams having this factorized structure. In this paper, we prove the first
property and assume that the second indeed holds, as has been shown to be the case elsewhere at eikonal level. 
We return to the issue of corrections to the above factorized form (at NE level) in Sec.~\ref{sec:low-theorem}.

To make these statements more direct, we begin by separating the path integral over the gauge boson
field into a product of integrals over hard and soft modes
\begin{equation}
\int\measure{A^\mu}\equiv\int\measure{A^\mu_s}\measure{A^\mu_h}
\label{Asep}
\end{equation}
The precise definition of $A^\mu_s$ and $A^\mu_h$ amounts to specifying a surface in the multi-boson momentum 
space that separates it into two distinct regions corresponding to soft and hard modes. Such a surface
is, in general, very complicated \cite{Sterman:1987aj}. Its precise definition does not concern us in what follows, 
where we characterize soft radiation by the fact that one may neglect the recoil of the eikonal particles\footnote{We will consider
corrections to this idea when discussing next-to-eikonal exponentiation in section \ref{sec:next-eikon-expon}.}.
However, the fact that such a surface exists allows us to introduce the factorization of Eq.~(\ref{Asep}), given
that the path integral is a product of integrals over gauge fields of definite momentum.

The above separation is not gauge invariant, which can be easily seen as follows. Consider a given
soft gauge field $A^\mu_s$, whose momentum modes live only in the soft region of momentum space. A general 
gauge transformation has the form
\begin{equation}
A^\mu_s(x)\rightarrow A^\mu_s(x)+\partial^\mu\xi(x),
\label{gauge}
\end{equation}
for some function $\xi(x)$. Transforming to momentum space, $\xi(x)$ may well have momentum modes defined in
the hard region of momentum space, and thus the transformed gauge field will have in general both soft and hard
components. Instead, both the soft and hard gauge fields obey a restricted gauge invariance given by the 
momentum space analogue of Eq.~(\ref{gauge})
\begin{equation}
A^\mu_{s,h}(k)\rightarrow A^\mu_{s,h}(k)+k^\mu\xi'(k).
\label{gauge2}
\end{equation}
Here $\xi'(k)$ is non-zero only if $k$ is in the soft and hard regions for $A^\mu_s$ and $A^\mu_h$ respectively.

We now formally define the hard interaction as
\begin{align}
H(x_1,\ldots,x_n)&=\int\measure{A^\mu_h}\,\measure{\phi}\,\measure{\phi^*}\frac{1}{i^n}\frac{\delta}{\delta J(y_1)}
\ldots\frac{\delta}{\delta J(y_n)}\matelem{y_1}{S-i\epsilon}{x_1}\ldots\matelem{y_n}{S-i\epsilon}{x_n}\notag
\\&\quad\times\exp\Big[iS[\phi,\phi^*,A^\mu]+i\int d^dx \big(J(x)\phi^*(x)+J(x)\phi^*(x)\big)\Big].
\label{Hdef}
\end{align}
This is analogous to the expression for a Green's function, except for the fact that the path integral
over soft gauge field modes $A^\mu_s$ has yet to be performed. Also, the factors $\matelem{y_i}{S-i\epsilon}{x_i}$ 
(i.e. inverse propagators for the particle in the background of the soft gauge field) truncate the external 
legs of the Green's function. 

We now define the further quantity
\begin{equation}
G(p_1,\ldots,p_n)=\int\measure{A^{\mu}_s}\,H(x_1,\ldots,x_n)\matelem{p_1}{(S-i\varepsilon)^{-1}}{x_1}\ldots
  \matelem{p_n}{(S-i\varepsilon)^{-1}}{x_n},
\label{eq:16}
\end{equation}
where a propagator factor has been associated with each external line, and the path integral over $A^\mu_s$
inserted. This latter integral does two things. Firstly, it generates all possible subgraphs within the
hard interaction $H$ (i.e. such that there are $n$ external lines emerging at 4-positions $\{x_i\}$), 
containing both soft and hard gauge boson modes.
Secondly, it produces soft radiation, both real and virtual, from the external lines. This is shown
schematically in figure \ref{Gdef}, where real and soft radiation is included in the soft blobs attached
to each external line. One thus sees that $G$ is a full Green's function of the theory, written in 
eikonally factorized form. Note that the propagator for the emitting particle in the presence of the background soft gauge field
is removed in Eq.(~\ref{Hdef}), and replaced in Eq.(~\ref{eq:16}) with the propagator sandwiched between states of
given initial position and final momentum.

To obtain the contribution to the scattering amplitude from the function $G(p_1,\ldots,p_n)$, 
one must truncate each external propagator. That is, for external leg $i$ one multiplies
by a factor $p_i^2+m^2$ to take account of the fact that the line is external and thus has 
no (free) propagator attached. In principle one must also divide by the residue of the
scalar propagator, arising from renormalization of the scalar field. However, 
this residue is unity due to the absence of self-interactions for the eikonal particle,
and also the fact that the gauge field is treated as a background. As is clear from 
Eq.~(\ref{eq:16}), one may treat each external line separately. Using the representation 
(\ref{eq:14}), one can rewrite each external line contribution as
\begin{align}
  \label{eq:18}
    i (p_f^2+m^2)&\matelem{p_f}{-i(S-i\varepsilon)^{-1}}{x_i}=i (p_f^2+m^2)\frac12 \int_0^\infty dT e^{-ip_f x_i-i\frac12(p_f^2+m^2-i\varepsilon)T} f(T) \notag\\
    &\quad=-e^{-i p_f x_i} \int_0^\infty dT \left(\frac{d}{dT}e^{-i\frac12(p_f^2+m^2)T}\right) \left(e^{-\frac12 \varepsilon T} f(T)\right)\notag\\
    &\quad=-e^{-i p_f x_i}\left(- f(0)- \int_0^\infty dT e^{-i\frac12(p_f^2+m^2)T}\left(\frac{d}{dT} e^{-\frac12\varepsilon T} f(T)\right)\right) \notag\\
    &\quad=-e^{-i p_f x_i}\left(- f(0)- \int_0^\infty dT
      e^{-i\frac12(p_f^2+m^2)T}\frac{d}{dT} f(T)\right)\,.
\end{align}
In the last step we have taken the limit $\varepsilon\rightarrow
0$. At this point one can let $p_f$ approach its mass shell and obtain
the simple result
\begin{equation}
  \label{eq:20}
  i(p_f^2+m^2)\matelem{p_f}{-i(S-i\varepsilon)^{-1}}{x_i}=e^{-ip_f x_i} f(\infty).
\end{equation}
The limit $T\rightarrow \infty$ of $f(T)$ in Eq.~\eqref{eq:20} allows us to simplify
the expression for $f$ in Eq.~\eqref{eq:13} by performing the Gaussian integral over
$p$. After shifting the integration variable $p\rightarrow p+A$ the result is
a path integral over $x$ only
\begin{equation}
  \label{eq:21}
  f(\infty)=\int_{x(0)=0} \measure{x}\, e^{i \int_0^\infty dt \left(\frac12 \dot{x}^2+(p_f+\dot{x})\cdot A(x_i+p_f t+x(t))+
      \frac i2\partial\cdot A(x_i+p_f t+x)\right)}.
\end{equation}
Thus, the eikonally factorized contribution to the scattering amplitude for a charged 
scalar in an abelian background field takes the form
\begin{equation}
  \label{eq:19}
  S(p_1, \ldots, p_n) = \int \measure{A^\mu_s} H(x_1, \ldots, x_n) 
\,  e^{-i p_1 x_1}f_1(\infty)\ldots e^{-i p_n x_n}f_n(\infty)\,e^{i S[A_s]}.
\end{equation}
with $f(\infty)$ given by Eq.~\eqref{eq:21}, and
the label of each $f(\infty)$ indicates the particular external
line. Also we have explicitly factored out the action for the soft gauge field,
which remains after the path integrals over the particle and hard
gauge fields. This form (\ref{eq:19}) will now enable us to find all-order expressions
for these amplitudes.

As a simple one-dimensional path integral, it can be further manipulated
using simple classical methods. The strictest approximation is to
neglect the fluctuations $x(t)$ and $p(t)$. This is equivalent to the eikonal
approximation in Feynman diagrams, and one sets $x=0,\; \dot{x}=0, \;
p=0$ and as well as neglecting the $\partial\cdot A$ and $A^2$
terms in Eq.~(\ref{eq:14}).  One then finds an Aharanov-Bohm-like phase
factor for the straight line trajectory
\begin{equation}
  f(\infty)\propto e^{i\int dx\cdot A(x)}.
\label{phase}
\end{equation}

Inserting this result into the path integral (\ref{eq:19}) where we
integrate over soft gauge field fluctuations the Wilson
lines, being linear in the soft gauge field $A^\mu_s$, 
act as a collection of classical source terms for the soft $A$-field, distributed along the classical trajectory.
\subsection{Eikonal exponentiation}
Now that we have established that in the eikonal approximation the
soft radiation is described by a Wilson line we can analyze what
happens in perturbation theory. Let us consider, without loss of generality,
an external line created at $x_i=0$ and in direction $n^\mu\equiv p_f^\mu$. 
Then Eq.~(\ref{phase}) becomes
\begin{equation}
  \exp\left[i\int_0^\infty dt n^\mu A_\mu(nt)\right].
\label{phase2}
\end{equation}
This can be written, after a Fourier transform to momentum space, as
 \begin{equation}
     i\int_0^\infty dt\, n^\mu A_\mu(nt)
     =-\int \frac{d^dk}{(2\pi)^d} \frac{n^\mu \tilde{A}_\mu(k)}{n\cdot k}.
 \end{equation}
Note that this is invariant under rescalings of the eikonal momentum $n^\mu$. 
As seen above, this acts as a source term for the soft gauge field when the
path integral over $A^\mu_s$ is performed. It can be
represented as a 1-photon vertex with the momentum space Feynman rule
\begin{equation}
      \includegraphics[scale=.6]{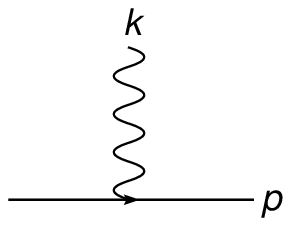} \qquad-\frac{n^\mu}{n\cdot k}\,,
\label{1gv}
\end{equation}
where momentum $k$ flows into the vertex. The path integral over the soft gauge field
generates all possible diagrams connecting numbers of source vertices. By the usual rules
of quantum field theory, one finds connected and disconnected diagrams, which in this case connect
the external lines given that the source vertices lie along the latter. The collection of all diagrams
exponentiates in terms of connected diagrams. 

\begin{figure}
\begin{center}
\scalebox{0.6}{\includegraphics{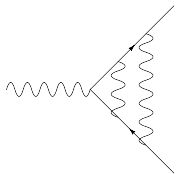}}
\caption{Example of a disconnected subdiagram between two outgoing external lines, 
to be compared with the connected subdiagrams of figure \ref{exp1}.}
\label{discon}
\end{center}
\end{figure}

To illustrate this further, consider the case of a hard interaction with two outgoing eikonal
lines, an example of which is shown in figure \ref{exp1}. There one sees a number of
connected subdiagrams connecting the external lines. One also finds disconnected diagrams,
such as that shown in figure \ref{discon}.
However, given the usual property of exponentiation of disconnected diagrams in 
quantum field theory, one has:
\begin{equation}
  \sum G=\exp\left[\sum G_c\right].
\end{equation}
The sum on the left is over all subdiagrams $G$, while the sum on the
right is over all connected diagrams $G_c$. 

We have thus succeeded in showing that the exponentiation of soft radiative corrections
in the eikonal limit can be related to the exponentiation of disconnected diagrams. We now 
consider what happens when next-to-eikonal corrections are considered.
\subsection{Next-to-eikonal exponentiation}
\label{sec:next-eikon-expon}
The analysis of the previous section relied on the fact that the factors $f(\infty)$ 
describing soft radiation from the external lines are written as path integrals in $x$.
This allowed the straightforward interpretation of the eikonal limit as the limit in
which the emitting particle follows a classical free path. However, this identification
also allows one to go easily beyond the eikonal approximation. If the emitted radiation
is soft but with non-negligible momentum, the classical path is still a good approximation to the 
equations of motion for the eikonal particle, and one can examine deviations from the 
straight-line path in a systematic expansion. Given that this does not affect the 
interpretation of gauge boson emission vertices in terms of disconnected subdiagrams,
one still expects such corrections to exponentiate.

To formalize this argument, we reconsider the external line given by Eq.~(\ref{eq:6}),
where again we take $x_i=0$ without loss of generality.
Given that the external eikonal particles have light-like momenta, one may write 
$p_j=\lambda n_j$, where $n_j^2=0$. Then Eq.~({\ref{eq:21}) becomes
\begin{multline}
\label{finfty2}
  f(\infty)=\int_{x(0)=0} \measure{x}\, \exp\Big[i\int_0^\infty dt 
\Big(\frac12\dot{x}^2+(\lambda n+\dot{x})\cdot A(\lambda nt+x)
      +\frac{i}{2}\partial\cdot A(\lambda n t+x)\Big)\Big].
\end{multline}
One now clearly sees that in the limit $\lambda\rightarrow\infty$, one may neglect
all terms involving $x$, $\dot{x}$ and $\partial\cdot A$, leaving precisely the eikonal
approximation discussed in the previous section. That is, fluctuations about the classical
free path are suppressed by inverse powers of $\lambda$. By expanding in $\lambda$,
one keeps the first subleading corrections to the eikonal approximation, i.e. 
corresponding to the next-to-eikonal (NE) limit.

For subsequent purposes it is more convenient to rescale the time variable 
$t\rightarrow t/\lambda$, so that Eq.~(\ref{finfty2}) becomes
\begin{multline}
\label{eq:22}
  f(\infty)=\int_{x(0)=0} \measure{x}\, \exp\Big[i\int_0^\infty dt 
\Big(\frac\lambda2\dot{x}^2+(n+\dot{x})\cdot A(nt+x)
      +\frac{i}{2\lambda}\partial\cdot A(nt+x)\Big)\Big].
\end{multline}
The first term in the exponent is now $\sim{\cal O}(\lambda)$, but gives rise to a 
propagator for $x(t)$ which is ${\cal O}(\lambda^{-1})$ by virtue of being the 
inverse of the quadratic operator in $x(t)$. The remaining terms generate effective
vertices for soft gauge boson emission in the NE limit, which one can again
interpret as source terms for the soft gauge field. Thus, following the reasoning in 
the previous section, one finds that these NE corrections exponentiate as before, i.e. one has:
\begin{equation}
  f(\infty)=\exp\left[\sum G_c^x\right]\,,
\end{equation}
where $G_c^x$ are connected (through $x$ propagators) diagrams along external lines, and located on the latter
by vertices derived by a systematic expansion of Eq.(\ref{eq:22}) in $\lambda^{-1}$.
At LO, one recovers the eikonal approximation of the previous section.
To obtain the NE approximation one must gather all terms $\sim{\cal O}(\lambda^{-1})$,
which can be described as follows.

Firstly, NE graphs must have at most one propagator for the emitting particle $x(t)$, due
to its being ${\cal O}(\lambda^{-1})$ as remarked above. There is also a NE vertex originating
from the term in $\partial\cdot A$, and a given NE graph containing such a vertex must then 
contain no propagator factors for $x(t)$. We examine in detail the NE Feynman rules that result
from Eq.~(\ref{eq:22}) in appendix \ref{sec:next-eikonal-feynman}, and show that they agree with 
the results one obtains in standard perturbation theory after expanding to NE order. The advantage
of the above representation, however, is that exponentiation of these corrections is manifest.
\subsection{Exponentiation for spinor particles}
We have so far only considered the case of a scalar eikonal particle. For emitting fermions (with a similar
expression for combinations of fermions and antifermions etc.), we 
write the definition of the hard interaction as
\begin{align}
H(x_1,\ldots,x_n)&=\int\measure{A^\mu_h}\,\measure{\psi}\,\measure{\bar{\psi}}\frac{1}{i^n}\frac{\delta}{\delta \bar{\eta}(y_1)}
\ldots\frac{\delta}{\delta \bar{\eta}(y_n)}\matelem{y_1}{S_0-i\epsilon}{x_1}\ldots\matelem{y_n}{S_0-i\epsilon}{x_n}\notag
\\&\quad \quad \quad\times\exp\Big[iS[\psi,\bar{\psi},A^\mu]+i\int d^dx \big(\bar{\eta}(x)\psi(x)+\bar{\psi}\eta(x)\big)\Big],
\label{Hdefferm}
\end{align}
where $S_0-i\epsilon$ is the free fermion inverse propagator. The eikonally factorized Green's
function has the same form as before (Eq.~(\ref{eq:16})), where the propagator in the presence of the 
background gauge field is given by Eqs.~(\ref{fermpath}). Truncating the external lines of
the full Green's function\footnote{In the spinor case we defined the evolution operator $U(T)$ as involving only
the denominator of Eq.~(\ref{fermprop}). The leftover factor in the numerator indeed combines correctly with the 
inverse free propagator to give a factor $p_f^2+m^2$ as in the scalar case of Eq.~(\ref{eq:18}).}, 
one finds that the eikonally factorized scattering amplitudes are given by the 
same expression Eq.~(\ref{eq:19}), but where the external line factor is now:
\begin{multline}
\label{fferm}
  f(\infty)=\int_{x(0)=0} \measure{x}\, \exp\Big[i\int_0^\infty dt 
\Big(\frac\lambda2\dot{x}^2+(n+\dot{x})\cdot A(nt+x)\\
      +\frac{i}{2\lambda}\partial\cdot A(nt+x)+\frac{1}{2\lambda}\sigma^{\mu\nu}F_{\mu\nu}\Big)\Big],
\end{multline}
and we have rescaled the time variable $t\rightarrow t/\lambda$ as before. The proof of exponentiation
up to NE order proceeds directly as in the scalar case, except for the additional magnetic moment vertex
which, although absent in the scalar case, does nothing to invalidate the proof.
Note that, due to suppression by $\lambda$, the additional vertex is indeed of NE order, as expected given that in the 
strict eikonal limit, radiation is insensitive to the spin of the emitting particle.

One may worry about ordering of Dirac matrices when the exponential of Eq.~(\ref{fferm}) is expanded. 
However, this is not an issue due to the fact that the magnetic moment vertex is NE and thus occurs in
each diagram only once at this order.

The preceding analysis has shown that in eikonally factorized Green's functions, soft gauge boson
corrections exponentiate up to NE order. This is not yet a proof that such corrections exponentiate
in matrix elements themselves, which include Green's functions not having 
an eikonally factorized structure. At strictly eikonal level (as is well known), one may in fact
ignore contributions from diagrams which are not eikonally factorized. At NE order, however, contributions
arise from diagrams in which a soft emission connects an external line with the hard interaction.
This is the subject of the following section.
\subsection{Low's theorem}
\label{sec:low-theorem}
In the previous section we have demonstrated
exponentiation for next-to-eikonal photon emissions
from external lines. That is, the exponentiation holds for scattering amplitudes
having the eikonally factorized form of figure \ref{Gdef}. However, at NE order
there are also corrections to the exponentiation arising from soft gluon emissions
which land on an external line, having originated from inside the hard interaction.
A given matrix element then has the schematic form (up to next-to-eikonal level):
\begin{equation}
{\cal M}=\exp\left[{\cal M}^{\text{E}}+{\cal M}^{\text{NE}}\right](1 + {\cal M}_r).
\label{Mstruc}
\end{equation}
Here ${\cal M}^{\text{E}, \text{NE}}$ collect the eikonal and next-to-eikonal diagrams
from eikonally factorized Green's functions respectively, and ${\cal M}_r$ is a remainder
term which does not exponentiate, and contains NE contributions from diagrams such as that shown
in figure \ref{remainder}.
\begin{figure}
\begin{center}
\scalebox{0.8}{\includegraphics{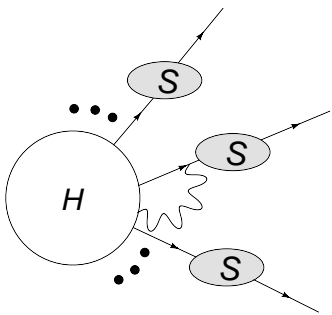}}
\caption{An example of a diagram which contributes to the remainder factor ${\cal M}_r$ of Eq.~(\ref{Mstruc}).
Such contributions are formed by taking an eikonally factorized Green's function, and adding a soft gluon emission
which lands on an external line, but originates from inside the hard interaction.}
\label{remainder}
\end{center}
\end{figure}
In what follows, we refer to emissions from within the hard interaction as {\it internal emissions}, and 
those originating from external lines as {\it external}.
Diagrams with internal emissions have been studied before in the literature. It has been shown for a fixed number of 
scalar external lines that, up to NE order, the sum of diagrams containing a soft emission (internal or external)
can be related to the scattering amplitude with no emissions~\cite{Low:1958sn}. This result is known as {\it Low's theorem}, and was
generalized to the case of spinor external lines in~\cite{Burnett:1967km}, an extension known as the 
{\it Low-Burnett-Kroll theorem}. Generalization to higher orders was considered in~\cite{DelDuca:1990gz}. The fact 
that graphs with an extra emission can be related simply to those without an emission means that, although the
remainder term does not have a formal exponential structure, it has an iterative form to all orders in perturbation theory.
In this section we discuss these properties in the path integral formalism adopted in this paper, in order to complete
our discussion of NE exponentiation. This also allows for a generalization of the ideas presented 
in~\cite{Low:1958sn,Burnett:1967km,DelDuca:1990gz}.

Our starting point is the expression for the $n$-particle scattering amplitude (see also Eq.~(\ref{eq:19}))
\begin{equation}
  \label{eq:19b}
  S(p_1, \ldots, p_n) = \int \measure{A_s} H(x_1, \ldots, x_n; A_s)
  e^{-i p_1 x_1}f(x_1,p_1;A_s)\ldots e^{-i p_n x_n}f(x_1,p_1;A_s)e^{iS[A_s]}\,,
\end{equation}
where we have explicitly indicated the dependence of the external leg factors (Eq.~(\ref{eq:22}))
on the soft gauge field. We also now consider the fact that the external lines are produced
at 4-positions $x_i\neq 0$ i.e.
\begin{multline}
  \label{eq:12}
  f(x_i,p_f,A_s) = \int_{x(0)=x_i}^{p(\infty)=p_f} \measure{x}\, \exp\Big[i\int_0^\infty dt
\Big(\frac\lambda2\dot{x}^2+q_i\,(n+\dot{x})\cdot A(x_i+nt+x)\\
      +\frac{i}{2\lambda}q_i\,\partial\cdot A(x_i+p t+x)\Big)\Big].
\end{multline}
We have also indicated the dependence of the hard part on the soft
photon field $A_s$ and rescaled this field so as to explicitly display the
dependence on the eikonal particle's electric charge $q_i$. Furthermore, we 
drop the subscript $s$ on the gauge
field in what follows, and leave implicit the path integrals over the external line
4-positions $x_i$.

Both $H$ and the $f$'s depend on the soft gauge field $A$. As discussed in section 
\ref{sec:soft-emiss-scatt}, the separation between soft and hard gauge modes leaves a 
residual gauge invariance, given by the usual form of Eq.~(\ref{gauge}), but where the 
function $\Lambda(x)$ has only soft modes when transformed to momentum space. It is implicitly
assumed that this in the case in the following. Under such a transformation, the external line
factors transform as
\begin{equation}
  \label{eq:37}
  f(x_i,p_f;A) \rightarrow f(x_i,p_f;A+\partial \Lambda) = e^{-iq \Lambda (x_i)} f(x_i,p_f;A),
\end{equation}
which follows from the definition of Eq.~(\ref{eq:20}).

In order for the path integral to remain invariant under the family of gauge transformations given by 
Eq.~(\ref{gauge}), the hard function $H$ must transform as
\begin{equation}
  \label{eq:38}
 H(x_1, \ldots, x_n; A) \rightarrow H(x_1, \ldots, x_n; A+\partial \Lambda)
 = H(x_1, \ldots, x_n; A) e^{i q_1 \Lambda (x_1) + \ldots + i q_n \Lambda (x_n)}.
\end{equation}
We can now use the gauge invariance to relate diagrams with soft emissions from inside
the hard interaction to similar diagrams with no emission, as follows.
First, one may expand both sides of Eq.~\eqref{eq:38} to first
order in $A$ and in $\Lambda$, which gives
\begin{multline}
  \label{eq:39}
  H(x_1, \ldots, x_n) + \int d^dx\, H^\mu(x_1, \ldots, x_n;x) (A_\mu(x) + \partial_\mu \Lambda(x))
 = \\
  H(x_1, \ldots, x_n) + \int d^dx\, H^\mu(x_1, \ldots, x_n;x) A_\mu(x) \\
 + i \int d^dx\, \Big(H(x_1, \ldots, x_n) \sum_j^n q_j \delta(x-x_j) \Big) \Lambda(x).
\end{multline}
When the path integral over the soft gauge field is performed, $H^\mu(x_1,\ldots,x_n;x)$ 
generates hard interactions which a single soft photon emission (with Lorentz index $\mu$).
Because $\Lambda(x)$ is arbitrary we infer
\begin{equation}
  \label{eq:40}
  -\partial_\mu H^\mu (x_1, \ldots, x_n;x) = i H(x_1, \ldots, x_n) \sum_j^n q_j \delta(x-x_j),
\end{equation}
where we have integrated by parts on the left hand side of Eq.~(\ref{eq:39}). In momentum space this 
becomes
\begin{equation}
  \label{eq:41}
    -k_\mu H^\mu (p_1, \ldots, p_n;k) =  \sum_j^n q_j  H(p_1, \ldots,p_j+k,\ldots, p_n).
\end{equation}
One can expand this up to first order in $k$ to obtain
\begin{equation}
  \label{eq:42}
    -k_\mu H^\mu (p_1, \ldots, p_n;k) =  \sum_j^n q_j k_\mu
    \frac{\partial}{\partial {p_j}_\mu} H(p_1, \ldots, p_n) \,,
\end{equation}
where the zeroth order term on the right hand side vanishes due to charge conservation
$\sum_j q_j=0$. Now, because $k^\mu$ is an arbitrary soft momentum, one may write
\begin{equation}
  \label{eq:43}
  H^\mu (p_1, \ldots, p_n;k) =  - \sum_j^n q_j \frac{\partial}{\partial {p_j}_\mu} H(p_1, \ldots, p_n) \,.
\end{equation}
This simply relates internal emission from the hard interaction with the same hard interaction
but with no emission. For a simple example of how this works using more traditional methods, we refer the
reader to appendix \ref{sec:low-example}.

In section \ref{sec:soft-emiss-scatt} we ignored the fact that the external lines emerge at positions
$x_i$ which are integrated over and thus in general non-zero. Taking this into account also leads to 
corrections of NE order (and beyond), which enter the above remainder term ${\cal M}_r$. To see this,
we write the eikonal one-photon source term as:
\begin{equation}
  \label{eq:44}
  -q \int \frac{d^dk}{(2\pi)^d} \frac{n\cdot A(k)}{n\cdot k} e^{ix\cdot k}.
\end{equation}
This can be expanded as
\begin{equation}
  \label{eq:45}
  -q \int \frac{d^dk}{(2\pi)^d}\, (1+ix\cdot k) \,\frac{n\cdot A(k)}{n\cdot k},
\end{equation}
where the first bracketed term corresponds to the eikonal approximation, and the 
second term (involving the factor $x\cdot k$) is suppressed by one power of momentum
and is thus a NE correction. We now combine these terms with the factors
of Eq.~(\ref{eq:43}), and up to NE order the scattering amplitude is given by a sum over
all such corrections, where each NE factor occurs at most once. One then finds
\begin{multline}
  \label{eq:46}
  S(p_1, \ldots, p_n) = \int \measure{A}
\Bigg[ \Big(- \sum_j^n q_j \frac{\partial}{\partial {p_j}_\mu} H(p_1, \ldots, p_n) \Big)
 \int \frac{d^dk}{(2\pi)^d} \, A_\mu(k) + \\
\int dx_1^d \ldots dx_n^d \,H(x_1, \ldots, x_n)
\Big(\sum_j q_j  \int \frac{d^dk}{(2\pi)^d}(-ix_j\cdot k) \frac{n\cdot A(k)}{n\cdot k} \Big)
e^{- i x_1\cdot p_1 - \ldots - i x_n \cdot p_n}\Bigg]\\
f (0,p_1;A) \ldots f (0,p_n;A)\,,
\end{multline}
where we have explicitly instated the integrals over the initial positions of the external lines
$\{x_i\}$. Performing these integrals, the scattering amplitude is given by
\begin{multline}
  \label{eq:47}
  S(p_1, \ldots, p_n) = \int \measure{A}
\Bigg[\int \frac{d^dk}{(2\pi)^d} \,\sum_j^n q_j
\Big(\frac{n_j^\mu}{n_j\cdot k} k_\nu \frac{\partial}{\partial {p_j}_\nu} -
  \frac{\partial}{\partial {p_j}_\mu}\Big) H(p_1, \ldots, p_n) A_\mu(k)\Bigg] \\ \times f (0,p_1;A) \ldots f (0,p_n;A)\,.
\end{multline}
Some comments are in order regarding the form and interpretation of this result. The external
line factors $f(0,p_i;A)$ contain exponentiated eikonal and NE terms, as discussed previously. 
Corrections to the NE exponentiation then arise due to the bracketed prefactor in Eq.~(\ref{eq:47}),
which contains a sum over different possible NE corrections. Such corrections contribute to the
remainder term ${\cal M}_r$ in Eq.~(\ref{Mstruc}), and do not exponentiate. However, Eq.~(\ref{eq:47})
shows that they can be obtained as derivatives of the hard interaction with no soft emissions. Thus,
the remainder term has an iterative structure to all orders in perturbation theory.

For scattering amplitudes, one may summarize this as follows. Leading eikonal logarithms arising from soft 
gluon emission exponentiate. NE logarithms do not exponentiate, but can be separated into the sum of a series
which does exponentiate, and a remainder sum which does not exponentiate, but is obtainable in principle to
all orders in the coupling constant.

To further clarify the above formulae, it is instructive to consider the case
where the hard interaction $H$ is a scalar. Then it can only depend
on Lorentz invariant products of momenta. The derivatives with respect to the 
4-momenta in Eq.~(\ref{eq:47}) can be reexpressed in terms of derivatives with respect to 
products of 4-vectors (corresponding to Mandelstam invariants in the hard scattering process).
One may verify that derivatives w.r.t. $p_i^2$ vanish, and in the case of two external lines ($n=2$)
there is only one scalar $p_1\cdot p_2$. Eq.~(\ref{eq:47}) then becomes
\begin{multline}
  \label{eq:47b}
  S(p_1, \ldots, p_n) = \int \measure{A}
\Bigg[\int \frac{d^dk}{(2\pi)^d} \,\Big(\frac{n_1^\mu(k\cdot n_2-k\cdot n_1)}{n_1\cdot k}+
\frac{n_2^\mu(k\cdot n_1-k\cdot n_2)}{n_2\cdot k}\Big)\\
\times\frac{\partial}{\partial (p_1\cdot p_2)} H(p_1, \ldots, p_n)
  A_\mu(k)\Bigg] f (0,p_1;A) \ldots f (0,p_n;A)\,.
\end{multline}
This is precisely the form one expects based on a conventional Feynman diagram treatment (see appendix
\ref{sec:low-example}), and one may interpret the bracketed factor in Eq.~(\ref{eq:47b}) as an extra
vertex describing soft emission from within the hard interaction.
\section{Non-abelian gauge theory}
\label{sec:non-abelian-gauge}
So far we have considered an abelian background gauge field. In the case where the
gauge field is non-abelian, the derivation of the scattering amplitude for eikonally factorized
diagrams proceeds similarly to Sec.~\ref{sec:soft-emiss-scatt}. Here we consider the simple case
of a hard interaction with two outgoing external lines. That is,
the analogue of Eq.~(\ref{eq:19}) can be written
\begin{equation}
  \label{Snonabel}
  S(p_1,p_2) = \int \measure{A^\mu_s} \,H^{i_1 i_2}(x_1,x_2) 
  e^{-i p_1 x_1}f_1^{i_1 j_1}(\infty)e^{-i p_2 x_2}f_2^{i_2 j_2}(\infty)e^{i S[A_s]}.
\end{equation}
Here $\{i_k\}$ and $\{j_k\}$ are indices in the fundamental representation of the gauge group,
such that the outgoing particles have indices $\{j_k\}$ and summation over repeated color indices is
implied. The external line factors $f^{i_kj_k}(\infty)$ have the form
\begin{equation}
  \label{fnonabel}
  f^{i_1 j_1}(\infty)=\left[\int_{x(0)=x_i} \measure{x}\, {\cal P}e^{i \int_0^\infty dt \left(\frac12 \dot{x}^2+(p_f+\dot{x})\cdot A(x_i+p_f t+x(t))+
      \frac i2\partial\cdot A(x_i+p_f t+x)\right)}\right]^{i_1 j_1},
\end{equation}
i.e. similar to before, but matrix-valued in color space due to the exponent being linear in 
the non-abelian gauge field $A^\mu=A^\mu_A t^A$, where $t^A$ is a generator of the gauge group. Furthermore,
there is a path ordering of the color matrices along the external line. As before, the external line 
factors act as source terms for the soft gauge field when the path integral over
$A^\mu_s$ is performed. However, it is no longer immediately clear that the soft corrections exponentiate. In the
abelian case, the exponentiation of soft gauge boson corrections was identified with the exponentiation of
disconnected diagrams between sources. Crucial to the combinatorics of this result is the fact that in the abelian case,
the source terms commute with each other. This is no longer true in the non-abelian case, due to the matrix valued nature
of the source terms, and also the path ordering of the exponential in Eq.~(\ref{fnonabel}). We will see, however, that
it is still possible to address exponentiation in the non-abelian case, by rephrasing the problem using the {\it replica trick}
of statistical physics (see appendix \ref{expproof} for another application i.e. the proof of
the exponentiation of disconnected diagrams in field theory). 
One can then write the scattering amplitude in a form such that extra structure emerges
in the exponent, whereby the contraction of soft gluon emissions between eikonal lines gives rise to a exponentiating
subset of diagrams. These can then be identified with the {\it webs} of~\cite{Frenkel:1984pz,Gatheral:1983cz}.

To simplify the discussion, we first restrict ourselves to the strict eikonal limit. Furthermore, we
consider the case of a hard interaction with the color singlet structure
\begin{equation}
H^{i_1i_2}(x_1,x_2)=H(x_1,x_2)\,\delta^{i_1i_2},
\label{Hsimp}
\end{equation}
where $\delta^{i_1i_2}$ is the Kronecker symbol. Such a structure arises in interactions where 
e.g. an incoming color singlet particle gives rise to the pair production of two hard final charged scalars
(the scalar analogue of $e^+ e^-$ pair production by a virtual photon), as shown in figure \ref{exp2}.
Given that, up to the NE corrections discussed in section \ref{sec:low-theorem}, one may consider the 
external lines as being created at $x=0$, one may take the hard interaction outside the path integral over $A_s$
in Eq.~(\ref{Snonabel}) to obtain (in this case)
\begin{equation}
  \label{Snonabel2}
  S(p_1, p_2) = H(p_1,p_2)\int \measure{A^\mu_s}\, f_1^{i j_1}(\infty)\,f_2^{i j_2}(\infty)\,e^{i S[A_s]}.
\end{equation}
Here $S[A_s]$ is the action for the soft gauge field which is independent of the emitting particles.
The product of external line factors, suppressing momentarily the color indices, is given by
\begin{equation}
f_1(\infty)f_2(\infty)=\left[{\cal P}e^{i \int dx_1\cdot A(x_1)}\right]
\left[{\cal P}e^{i \int dx_2\cdot A(x_2)}\right].
\label{fprod}
\end{equation}
The first factor is a Wilson line parametrized by $x_1(s)$, where $s=-t$ increases along the direction of the charge flow,
with $-\infty<s<0$. The second factor is a Wilson line parametrized by $x_2(s)$, with $s=t$ and $0<s<\infty$. One may
combine these into a single curve given by
\begin{equation}
x(s)=\left\{\begin{array}{l}x_1(s), \quad -\infty<s<0;\\x_2(s), \quad \quad0\leq s<\infty.\end{array}\right.
\label{willine2}
\end{equation}
Due to the path ordering in the definition of the Wilson line, one has the property
\begin{equation}
\left[{\cal P}e^{i \int dx_1\cdot A(x_1)}\right]\left[{\cal P}e^{i \int dx_2\cdot A(x_2)}\right]
={\cal P}e^{i \int dx\cdot A(x)},
\label{fprod2}
\end{equation}
so that, for the simple interaction considered here, one may combine the two external line factors into the single factor
\begin{equation}
f(\infty)=f_1(\infty)f_2(\infty)={\cal P}e^{i \int dx\cdot A(x)}.
\label{fprod3}
\end{equation}
The scattering amplitude of Eq.~(\ref{Snonabel2}) is now given by:
\begin{equation}
  \label{Snonabel3}
  S(p_1, p_2) = H(p_1,p_2)\,{\cal F},
\end{equation}
where
\begin{equation}
\label{fdef}
{\cal F}=\int \measure{A^\mu_s} f(\infty)e^{i S[A_s]}.
\end{equation}
We now consider the quantity
\begin{equation}
\label{FN}
{\cal F}^N=\left[\int\measure{A^\mu_1} f^{(1)}(\infty)e^{i S[A_1]}\right]\ldots
\left[\int\measure{A^\mu_N} f^{(N)}(\infty)e^{i S[A_N]}\right].
\end{equation}
Here $\{A^\mu_i\}$ are $N$ replicas of the soft gauge field (we have dropped the subscript $s$ for brevity),
and $S[A_i]$ the action for the $i^{th}$ replica. One has a different external line factor $f^{(i)}$ for
each replica field. Combining the path integrals using 
$\measure A^\mu\equiv\prod_i\measure A^\mu_i$, one can rewrite Eq.~(\ref{FN}) as 
\begin{equation}
\label{FN2}
{\cal F}^N=\int\measure{A^\mu} f^{(1)}(\infty)\ldots f^{(N)}(\infty)e^{i S[A_1]+\ldots+iS[A_N]}.
\end{equation}
The physical interpretation of this quantity is as follows. The external line factors, as in the abelian case,
contain sources for the gauge field. In this case, they generate diagrams containing any mixture
of the $N$ replica gauge fields, which span the external lines of the hard interaction (which in this case have become
a single external line). Each of the vertices for the emission of a gauge field replica has a non-trivial 
color structure, such that neither the external factors $f^{(i)}$ nor the vertices they give rise to commute.
However, by definition one has:
\begin{equation}
{\cal F}^N=1+N\log({\cal F})+{\cal O}(N^2).
\label{logF}
\end{equation}
It follows that, if one can extract a term in Eq.~(\ref{FN2}) that is linear in the number $N$ of replica fields, 
one has
\begin{equation}
{\cal F}=\exp\left[\sum W\right],
\label{Fexp}
\end{equation}
where the sum is over all diagrams $W$ that contribute at ${\cal O}(N)$. Crucially, we will find that 
not all diagrams in the theory have terms of ${\cal O}(N)$, so one recovers the property
of exponentiation of soft radiative corrections in terms of a subset of diagrams with certain properties.
The diagrams in this case will still contain replica fields. However, given that the gauge group of the replicated 
theory is the same as that in the standard theory, it must be true that the color structures 
of the subdiagrams which exponentiate are the same in the two theories. 

We now describe how to isolate the term linear in $N$ in Eq.~(\ref{FN2}). The product of external line
factors has the form:
\begin{equation}
f^{(1)}(\infty)\ldots f^{(N)}(\infty)={\cal P}\exp\left[\int dx \cdot A_1(x)\right]\ldots{\cal P}\exp\left[\int dx \cdot A_N(x)\right].
\label{fprod4}
\end{equation}
Ideally we want to write this as a single path-ordered exponential, so that one can identify the usual
rules of perturbation theory. This can be achieved by writing Eq.~(\ref{fprod4}) in the following form:
\begin{equation}
\prod_{i=1}^N{\cal P}\exp\left[\int dx \cdot A_i(x)\right]={\cal R}{\cal P}\exp\left[\sum_{i=1}^N\int dx\cdot A_i(x)\right],
\label{prod1}
\end{equation}
where we have introduced the {\it replica ordering} operator ${\cal R}$, defined such that
\begin{equation}
{\cal R}[A_i(x)A_j(y)]=\left\{\begin{array}{c}A_i(x)A_j(y),\quad i\leq j\\ A_j(y)A_i(x),\quad i>j\end{array}\right.,
\label{Rdef}
\end{equation}
with obvious generalization to higher numbers of operators. That is, ${\cal R}$ orders any product of matrix-valued fields
into a sequence of increasing replica number. Note that the resulting product is no longer strictly time ordered,
although the matrix fields of any given replica number remain time ordered. 

As before (and by analogy with conventional Feynman perturbation theory for non-Abelian gauge fields), the single 
exponent in Eq.~(\ref{Rdef}) acts as a collection of sources for the soft gauge field. The path integration over 
the soft gauge field generates diagrams containing
multiple replica emissions along the eikonal line, where the replica numbers are not necessarily ordered along 
the line (see figure \ref{replicaline}). However, the expression for a given diagram, as dictated by the source terms arising from
Eq.~(\ref{prod1}), involves replica ordered products of operators, each of which involves a color matrix. Thus, the 
color structure associated with each diagram is not the same as that which would result from conventional perturbation
theory, but rather that associated with the given replica-ordered product of matrix-valued fields. The subset of diagrams which
exponentiates then has a modified color structure, as is known to be the case for webs
~\cite{Gatheral:1983cz,Frenkel:1984pz,Sterman:1981jc}. 
\begin{figure}
\begin{center}
\scalebox{1.1}{\includegraphics{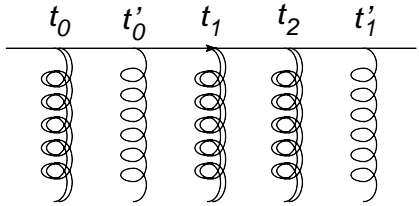}}
\caption{Example of an radiative corrections generated by the source terms of Eq.~(\ref{Rdef}), where two gauge boson replicas are shown.
Each set of replica emissions is time ordered, such that (in this case) $t_2>t_1>t_0$ and $t'_1>t'_0$.}
\label{replicaline}
\end{center}
\end{figure}
To see which diagrams $W$ actually contribute in Eq.~(\ref{Fexp}), one must consider contracting gluons
emitted from two or more vertices. Given that the gauge field replicas do not interact with each other
(i.e. are only tangled through color structure), one can clearly only contract gluons which have the 
same replica number $i$ and adjoint color index $A$. Here we consider this up to ${\cal O}(A_\mu^4)$ in the 
scattering amplitude (i.e. up to two gluon lines). Firstly, we need only consider contributions from vertices on 
different segments of the combined Wilson line $x(t)$, as those on the same segment ultimately 
give contributions proportional to (at least in covariant gauges) $p_1^2=0$ or $p_2^2=0$. 
Also, each diagram has a multitude of similar diagrams
obtained by permuting the replica labels. The operator ${\cal R}$ for each diagram orders the color matrices in the 
form
\begin{equation}
[t_1^{A_1}\ldots t_1^{A_{n_1}}]\ldots [t_N^{B_1}\ldots t_N^{B_{n_N}}]
\label{prodcols}
\end{equation}
i.e. a product of strings of color matrices, with one string for each replica (if present), and $n_i$ matrices for 
replica $i$. Given that the replicas
do not interact with each other, the color indices of each string in Eq.~(\ref{prodcols}) are contracted independently
of the other strings. That is, the color factor for each diagram has the form
\begin{equation}
\prod_{i=1}^N K_i^{A_1\ldots A_{n_i}}[t_1^{A_1}\ldots t_1^{A_{n_i}}],
\label{colfact}
\end{equation}
where $K_i^{A_1\ldots A_{n_i}}$ is a combination of factors involving $f^{ABC}$ and $\delta^{AB}$ which implements 
the color contractions for replica $i$ in the diagram being considered. Each of the strings of color matrices 
corresponding to a given replica number in Eq.~(\ref{colfact}) has
two color indices in the fundamental representation, thus by Schur's Lemma must be proportional to the 
identity. Hence, the color factor of a complete diagram in the replica ordered perturbation theory is the product of the
individual color factors associated with the subdiagrams formed from each replica separately. Furthermore, the ordering of
the factors in Eq.~(\ref{prodcols}) is unimportant, so that the color factors associated with the set of diagrams obtained from
a given diagram merely by permuting replica numbers are the same. There are $\!_NP_m=N!/(N-m)!$ such permutations, where $m$ is the
number of different replica species present in the diagram.

For one gluon emission, there is only one possible diagram, shown in figure \ref{nonabelfig}(a).
\begin{figure}
\begin{center}
\scalebox{0.9}{\includegraphics{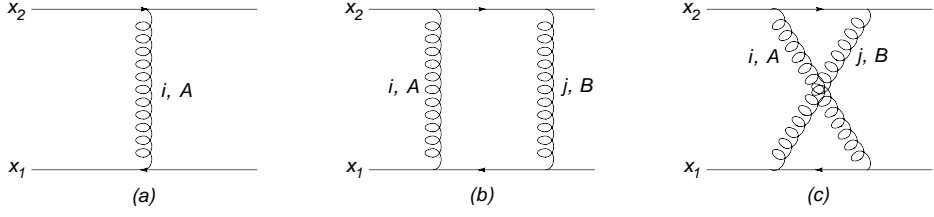}}
\caption{Diagrams that potentially contribute to the exponentiated contribution to the scattering amplitude in the case of two external lines connected 
by a color singlet structure, for one and two gluon emissions. In fact, only (a) and (c) contribute as discussed in the text.}
\label{nonabelfig}
\end{center}
\end{figure}
There is a sum over the replica number of the exchanged gluon, so that 
this diagram is clearly proportional to $N$, denoting a color structure that exponentiates. The color factor of this diagram
is
\begin{equation}
t^At^A=C_F,
\label{col1}
\end{equation}
where $t^A$ is a generator in the fundamental representation of the gauge group, and $C_F$ the relevant Casimir invariant. Note
that this is the same as the color factor in conventional perturbation theory, although things become more complicated
when more than one gluon is involved.

For two gluon emission, one has the two diagrams shown in Fig.~\ref{nonabelfig}(b,c). For each of these, one must consider
separately the cases where $i=j$ and $i\neq j$, because of the fact that the replica ordering operator ${\cal R}$ acts
differently in the two cases. 

For Fig.~\ref{nonabelfig}(b), in the case where $i=j$ one has a color factor
\begin{equation}
t_j^B\, t_i^A\, t_i^A\, t_j^B = C_F^2,
\label{colii}
\end{equation}
where we have explicitly indicated which color matrix is associated with each replica. 
When $i\neq j$, the color matrices in Eq.~(\ref{colij}) get reordered by the ${\cal R}$ operator i.e. one has
\begin{equation}
t_i^A\, t_i^A\, t_j^B\, t_j^B = C_F^2
\label{colij}
\end{equation}
for $i<j$, with a similar expression for $i>j$ (but where $i$ and $j$ are interchanged). 
The color factors of these diagrams are the same, and thus one may
combine the results for $i=j$ and $i\neq j$. Then one sees that the contribution from Fig.~(\ref{nonabelfig})(b) is 
${\cal O}(N^2)$.

For Fig.~\ref{nonabelfig}(c), the $i=j$ case has the color factor
\begin{equation}
t_i^A\, t_j^B\, t_i^A\, t_j^B=C_F^2-\frac{C_FC_A}{2}.
\label{colii2}
\end{equation}
When $i\neq j$ one has
\begin{equation}
t_i^A\, t_i^A\, t_j^B\, t_j^B=C_F^2.
\label{colij2}
\end{equation}
Note that the color factors for $i=j$ and $i\neq j$ are now different, such that the two cases do not combine to give a term
${\cal O}(N^2)$. There are $N$ diagrams where $i=j$, and $\!_NP_2=N(N-1)$ diagrams where $i\neq j$. 
Thus the term linear in $N$ has a color factor:
\begin{equation}
N\left(C_F^2-\frac{C_FC_A}{2}\right)+(-N)C_F^2=N\left(-\frac{C_FC_A}{2}\right).
\end{equation}

The above discussion can be summarized as follows. Up to two gluon emissions, a subset of diagrams exponentiates. Namely, the one gluon
emission diagram of Fig.~\ref{nonabelfig}(a), and the crossed gluon diagram of Fig.~\ref{nonabelfig}(c). Fig.~\ref{nonabelfig}(b) does 
not contribute due to being ${\cal O}(N^2)$, and Fig.~\ref{nonabelfig}(c) has a color factor which differs from that of conventional
perturbation theory, and indeed is precisely the modified color factor associated with the known webs of 
\cite{Gatheral:1983cz,Frenkel:1984pz,Sterman:1981jc}.

Note at this order in $\alpha_S$ one also has diagrams containing gluon self-interactions, of which there are two possibilities, shown
in Fig.~\ref{nonabelfig2}.
\begin{figure}
\begin{center}
\scalebox{0.9}{\includegraphics{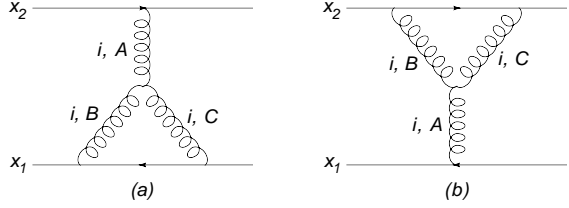}}
\caption{Diagrams contributing to the exponentiated scattering amplitude at ${\cal O}(\alpha_S^4)$, and involving the 3-gluon vertex.}
\label{nonabelfig2}
\end{center}
\end{figure}
There are $N$ such diagrams in each case, so that they also exponentiate (i.e. are webs) with a color factor equal to the ordinary one.

We now consider the generalization of the above remarks to higher orders in perturbation theory. The subdiagrams which 
exponentiate can be characterized by the fact that they are two-eikonal irreducible. That is, one cannot disconnect each
diagram by cutting eikonal lines in two places. This property is well-known~\cite{Frenkel:1984pz,Gatheral:1983cz}, but
we prove it in the following using the methods outlined above.

Consider the general two-eikonal reducible diagram shown in Fig.~\ref{twoeikred} (where each of the subdiagrams could
itself be reducible). 
\begin{figure}
\begin{center}
\scalebox{0.9}{\includegraphics{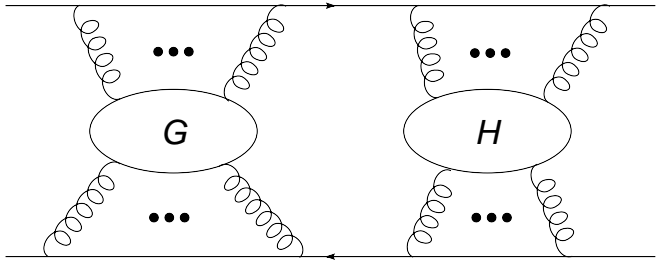}}
\caption{A two-eikonal reducible diagram, in the simplest case of two disconnected components with respect to the external lines.}
\label{twoeikred}
\end{center}
\end{figure}
We focus on a given gauge boson replica $i$ in subdiagram $H$ in Fig.~\ref{twoeikred}, and consider first the case where subdiagram $G$ contains
no gauge bosons with replica number $i$. According to the Feynman rules of the replica-ordered perturbation theory described above,
the diagram of Fig.~\ref{twoeikred} then gives a color factor
\begin{equation}
K_i^{A_1\ldots A_n\,B_1\ldots B_m}\left[t_H^{A_1}\ldots t_H^{A_n}\right]\left[t_H^{B_m}\ldots t_H^{B_1}\right]\times \prod_{j\neq i} C(j),
\label{tprodij}
\end{equation}
where $C(j)$ is the color factor associated with replica $j$, and $K_i^{A_1\ldots A_n\,B_1\ldots B_m}$ is the color contraction 
factor for replica $i$ introduced above. That is, $\prod C(j)$ contains the results of all contractions in $G$ as well as those of $H$ 
that do not involve replica number $i$. The indices $\{A_k\}$ and $\{B_k\}$
denote emissions from the lower and upper eikonal lines respectively. Now consider the case where $G$ as well as $H$
contains emissions with replica number $i$. Then (due to the reducible structure) one can write $K_i$ as a product of factors
$S_{G_i}$, $S_{H_i}$ for each subdiagram, so that the color factor associated with Fig.~\ref{twoeikred} is
\begin{align}
&K^{A_{n_H+1}\ldots A_{n}\,B_{m_H+1}\ldots B_{m}}_{G_i}\,S^{A_1\ldots A_{n_H}\,B_1\ldots B_{m_H}}_{H_i}\left[t_H^{A_1}\ldots t_H^{A_{n_H}}\right]\notag\\
&\times\quad\left[t_G^{A_{n_H+1}}\ldots t_G^{A_n}\right]\left[t_G^{A_{m}}\ldots t_G^{A_{m_H+1}}\right]
\left[t_H^{B_{m_H}}\ldots t_H^{B_1}\right]\times \prod_{j\neq i} C(j).
\label{tprodii}
\end{align}
We may use the fact that
\begin{equation}
K^{A_{n_H+1}\ldots A_{n}\,B_{m_H+1}\ldots B_{m}}_{G_i}\left[t_G^{A_{n_H+1}}\ldots t_G^{A_n}\right]\left[t_G^{A_{m}}\ldots t_G^{A_{m_H+1}}\right]\propto I,
\end{equation}
where $I$ is the identity in color space. This follows given that the contribution in color space from a given replica in a disconnected subdiagram $G$
has two indices in the fundamental representation. By Schur's Lemma, this must be proportional to the identity i.e. the 
only possible two-index invariant tensor. We may then rewrite Eq.~(\ref{tprodii}) as 
\begin{align}
&K^{A_{n_H+1}\ldots A_{n}\,B_{m_H+1}\ldots B_{m}}_{G_i}\,K^{A_1\ldots A_{n_H}\,B_1\ldots B_{m_H}}_{H_i}\left[t_H^{A_1}\ldots t_H^{A_{n_H}}\right]\notag\\
&\times\quad\left[t_H^{B_{m_H}}\ldots t_H^{B_1}\right]\left[t_G^{A_{n_H+1}}\ldots t_G^{A_n}\right]\left[t_G^{A_{m}}\ldots t_G^{A_{m_H+1}}\right]
\times \prod_{j\neq i} C(j).
\label{tprodii2}
\end{align}
This has the same color structure as would arise if one were considering a different replica number in $G$ than has been considered in $H$. 
I.e. one can absorb all $G$-dependent factors into the product $\prod C(j)$, such that Eq.~(\ref{tprodii2}) 
is then the same as Eq.~(\ref{tprodij}). The color structures
of the subdiagrams are thus independent, and it follows that the number of ways of forming the total diagram in Fig~\ref{twoeikred} is the product
of the number of ways of forming the individual subdiagrams. For each subdiagram this is at least $\propto N$, such that the contribution from
diagrams which are two-eikonal reducible is at least $\propto N^2$ (in general, $\propto N^M$, where $M$ is the number of disconnected subdiagrams).
Thus, two-eikonal reducible diagrams do not exponentiate.

In fact, one can proceed further and obtain a general solution for the color factors of the diagrams to all orders. Consider a given diagram 
consisting of $n_c$ connected pieces (i.e. gluons connected by self interactions or fermion bubbles). For each such diagram, we then consider
the set $\{P\}$ of {\it partitions}. These are sets containing a number $n(P)$ of subgraphs $g$, each of which
contains only one replica (see Fig. \ref{partitions}). Permuting the replica numbers (but keeping the subgraphs $g$ intact) corresponds to the same partition 
(see Fig.~\ref{partitions2}), such
that there are $\!_N P_{n(P)}$ distinct diagrams in each partition, each of which has the same color factor
\begin{equation}
\prod_{g\in P} C(g),
\label{gcol}
\end{equation}
where $C(g)$ is the color factor associated with subgraph $g$. The total color factor of the complete diagram is now given by
\begin{equation}
\sum_P \!_N P_{n(P)}\prod_{g\in P} C(g)
\label{Pcolsum}
\end{equation}
i.e. one sums over all possible partitions, each of which has a color factor given by Eq.~(\ref{gcol}), and weighted by the number of diagrams
represented by each partition. The only dependence on the replica number resides in the factor $\!_N P_{n(P)}$, which has the form
\begin{equation}
\!_N P_{n(P)}= (-1)^{n(P)-1}(n(P)-1)! N+{\cal O}(N^2).
\label{permexp}
\end{equation}
The contribution from a given complete diagram $G$ which is linear in $N$ is thus given by
\begin{equation}
\bar{C}(G)=\sum_P(-1)^{n(P)-1}(n(P)-1)!\prod_{g\in P} C(g).
\label{modcolsol}
\end{equation}
The bar on the left hand side denotes the fact that this is the modified color factor associated with $G$, rather than the
color factor one would obtain in conventional perturbation theory (i.e. without the ${\cal R}$ operator). The color factors 
on the right hand side, $C(g)$, are the ordinary color factors associated with each replica subgraph $g$.
\begin{figure}
\begin{center}
\scalebox{1.0}{\includegraphics{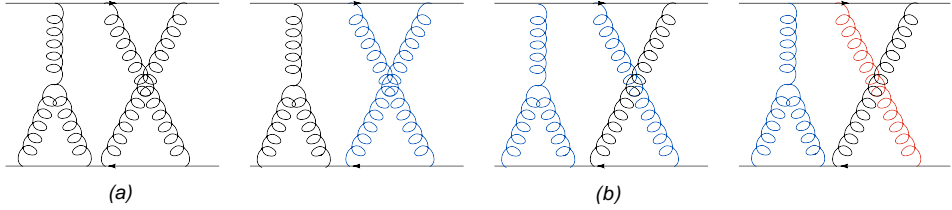}}
\caption{Examples of (a) a diagram containing 3 connected pieces; (b) three possible partitions generated by this diagram, where colors
represent distinct replica numbers. Permuting colors gives the same partition.}
\label{partitions}
\end{center}
\end{figure}
\begin{figure}
\begin{center}
\scalebox{1.0}{\includegraphics{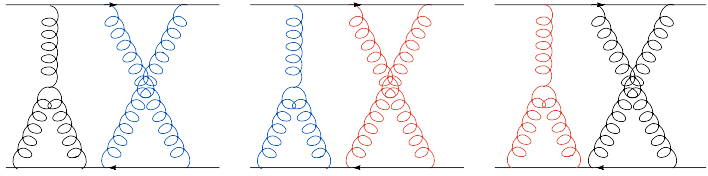}}
\caption{Examples of diagrams in the same partition, where colors represent distinct replica numbers.}
\label{partitions2}
\end{center}
\end{figure}
Eq.~(\ref{modcolsol}) is a closed form solution for the modified color factor associated with any given diagram, in that modified
color factors only appear on the left-hand side. It has the property,
as it must do from the above considerations, of being zero if $G$ is not a web. As a simple example (already encountered above),
consider the diagram of Fig.~\ref{nonabelfig}(c). There are two possible partitions. Either the two gluons have the same replica number, 
i.e. one subgraph with color factor $C(\figX)$, or they have different color factors i.e. two subgraphs with total color factor $C(\figI)C(\figI)$.
Then Eq.~(\ref{modcolsol}) gives a total modified color factor
\begin{equation}
\bar{C}(\figX)=C(\figX)-C(\figI)C(\figI),
\label{CX}
\end{equation}
which gives $-\frac12 C_FC_A$ as required.

To further clarify the discussion, we consider here a pair of three loop examples. We consider first the diagram represented in the
above shorthand notation by $\figLtwo$. From Eq.~(\ref{modcolsol}), one has schematically
\begin{align}
\bar{C}(\figLtwo)&=C(\figLtwo)-2C(\figI)C(\figX)-C(\figI)C(\figII)+2C(\figI)^3\notag\\
&=C(\figLtwo)-2\bar{C}(\figI)\bar{C}(\figX)-\bar{C}(\figI)^3,
\label{col3one}
\end{align}
where in the second line we have used the fact that $\bar{C}(\figI)=C(\figI)$ and $\bar{C}(\figX)=C(\figX)-C(\figI)^2$. 
The second line then agrees with the modified color factor given in \cite{Gatheral:1983cz,Frenkel:1984pz}. 

Our second example is $\figIX$, for which one has
\begin{align}
\bar{C}(\figIX)&=C(\figIX)-C(\figI)C(\figX).
\label{col3two}
\end{align}
Reducibility implies $C(\figIX)=C(\figI)C(\figX)$, and thus $\bar{C}(\figIX)=0$, as expected for a non-web.

To summarize, the above replica trick has allowed us to determine the subset of diagrams which exponentiate in non-abelian theory i.e. the diagrams
which exponentiate are those which have a term linear in the number of replicas $N$. This is related to the original gauge theory 
(with no replicas) as follows. Firstly, the replica gluons all have the same self-interactions and scalar-gluon interactions, but do not 
interact with each other. Thus in any diagram one can replace replica gluons with original gluons to yield the same kinematic result.
Then the color weights of the exponentiating diagrams are precisely those found above i.e. modified with respect to the original theory.
The structure of two-eikonal irreducible diagrams with modified color factors is precisely that of web exponentiation, described in 
\cite{Frenkel:1984pz,Berger:2003zh}.

The above discussion proceeds similarly in the case of fermionic emitting particles, given that the gauge boson emission 
vertices which distinguish scalar from fermion emitters only appear at NE order. We discuss subleading corrections in the next section.
\subsection{Non-abelian exponentiation at NE order}
The significance of the above derivation of web exponentiation in terms of the path integral method is that one may then
easily extend the analysis to NE order, using the $\lambda$-scaling technique discussed in section \ref{sec:next-eikon-expon}.
We have already shown that a subset of NE corrections exponentiate in the abelian case, and that eikonal corrections in the
non-abelian case exponentiate. Thus, it is not surprising that a subset of non-abelian NE terms exponentiates.

In this section we again consider the simple case of two external lines emerging from a hard interaction which has a 
color singlet structure. Then the scattering amplitude factorizes as in Eq.~(\ref{fdef}), but where the external line
factor is given by
\begin{multline}
\label{fferm-nonabel}
  f(\infty)=\int_{x(0)=0} \measure{x}\,{\cal P} \exp\Big[i\int_0^\infty dt 
\Big(\frac\lambda2\dot{x}^2+(n+\dot{x})\cdot A(nt+x)\\
      +\frac{i}{2\lambda}\partial\cdot A(nt+x)-\frac{1}{2\lambda}\sigma^{\mu\nu}F_{\mu\nu}\Big)\Big],
\end{multline}
One next performs the path integral in $x$ (as detailed for the abelian case in appendix \ref{sec:next-eikonal-feynman}),
which gives
\begin{align}
\label{fferm-nonabel2}
  f(\infty)=&{\cal P}\exp\Big[i\int_0^\infty dt\, n\cdot A(nt)-\frac{1}{2\lambda}\int_0^\infty dt\, \partial\cdot A(nt)\notag\\
&\quad+\frac{1}{2}\int_0^\infty dt\,\int_0^\infty dt'\dot\langle{x}^\mu(t)\dot{x}^\nu(t')\rangle A_\mu(nt)A_\nu(nt')\notag\\
&\quad-\int_0^\infty dt\int_0^\infty dt'\,\langle\dot{x}^\mu(t) x^\alpha(t')\rangle n^\nu A_\mu(nt)\partial_\alpha A_\nu(nt')\notag\\
&\quad-\frac{1}{2}\int_0^\infty dt\int_0^\infty dt'\,\langle x^\alpha(t) x^\beta(t')\rangle n^\mu n^\nu \partial_\alpha A_\mu(nt)\partial_\beta A_\nu(nt')\notag\\
&\quad+\frac{i}{2}\int_0^\infty dt\, n^\mu \langle x^\nu(t)x^\alpha(t)\rangle\partial_\nu\partial_\alpha A_\mu(nt)\Big].
\end{align}
For reasons that will become clear, we have stayed in position space in the exponent. Inserting the correlators of the $x$
fields (see Eqs.~(\ref{correlators})), Eq.~(\ref{fferm-nonabel2}) becomes:
\begin{align}
\label{fferm-nonabel3}
  f(\infty)=&{\cal P}\exp\Big[i\int_0^\infty dt\, n\cdot A^A(nt)\,t^A-\frac{1}{2\lambda}\int_0^\infty dt\, \partial\cdot A^A(nt)t^A
\notag\\
&\quad+\frac{1}{\lambda}\int_0^\infty dt\,\eta^{\mu\nu}A^A_\mu(nt)\,A^B_\nu(nt)\frac12\{t^A,t^B\}
+\frac{i}{2\lambda}\int_0^\infty dt \,n^\mu \partial^\nu\partial_\nu A^A_\mu(nt)\,t^A\notag\\
&\quad-\frac{1}{\lambda}\int_0^\infty dt\int_0^t dt'\,\eta^{\mu\alpha} n^\nu A^A_\mu(nt)\partial_\alpha\, A^B_\nu(nt')\,t^A\,t^B\notag\\
&\quad-\frac{1}{\lambda}\int_0^\infty dt\int_0^t dt'\, t' n^\mu n^\nu \partial_\alpha A^A_\mu(nt)\,\partial_\beta \,A^B_\nu(nt')\,t^A\, t^B
\Big].
\end{align}
Here we have explicitly factored out color matrices from the non-abelian gauge fields. Note that path ordering (i.e. time ordering
in this case) has appeared in some of the terms in the exponent, due to the $\Theta$ function occurring in the correlators of 
Eq.~(\ref{correlators}). There are two types of vertex occurring in Eq.~(\ref{fferm-nonabel3}) - those that depend on a single time
(in the first two lines), and those that depend on two different times (in the third and fourth lines). Of the former type, there are one gluon 
vertices and a two gluon vertex. For these vertices, the 
arguments of the previous section (involving the replica trick) carry forward with minimal modification, 
and one has eikonal exponentiation up to NE webs, where each NE vertex must occur only once per diagram. 
The additional webs that appear at two gluon order are shown in figure \ref{nonabelfig3}. At higher orders, the NE webs again have the property
of being two-external-line irreducible. 
\begin{figure}
\begin{center}
\scalebox{0.9}{\includegraphics{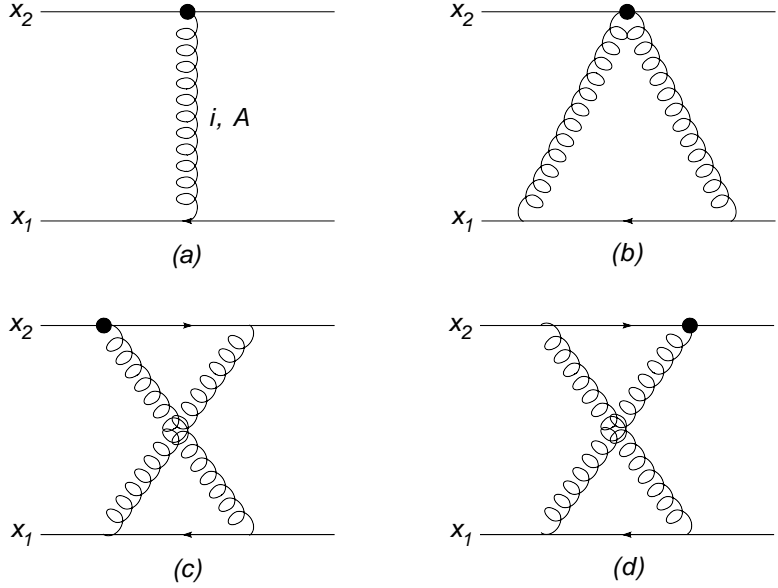}}
\caption{Webs involving local vertices occurring at NE order, where $\bullet$ represents a vertex of NE order. Additional diagrams 
arise by reflecting each diagram shown about the horizontal axis.}
\label{nonabelfig3}
\end{center}
\end{figure}
The time ordered vertices in the third line of Eq.~(\ref{fferm-nonabel3}) generate diagrams such as those shown in figure
\ref{correl}, which involve correlated emissions from different positions on the external line, but where additional eikonal emissions
may occur in between. One must then sum over all possible correlations (pairs of gluons). 
There are two types of such diagrams. Firstly, diagrams whose structure is such that they form a web at
eikonal level (i.e. are two-eikonal irreducible in that case). The time ordered vertices then implement correlations between 
pairs of gluons in the same web. Secondly, one has diagrams that would be two-eikonal reducible at eikonal level (i.e. are a product
of webs), but which become irreducible at NE level due to correlations between gluons in separate webs (e.g. Fig.~\ref{correl}). The sum over
all such diagrams then enters the exponent of Eq.~(\ref{Fexp}).
\begin{figure}
\begin{center}
\scalebox{0.9}{\includegraphics{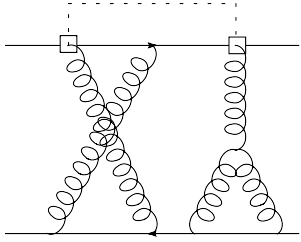}}
\caption{Example of a web involving the NE vertex (denoted here by $\Box$), which correlates gluons emitted from different positions on the external line.
As shown here, further eikonal emissions may occur between the correlated NE emissions.}
\label{correl}
\end{center}
\end{figure}
To summarize, the set of NE corrections resulting from non-abelian gauge boson emission outside of the hard interaction exponentiate.
The exponent involves a subset of diagrams, NE webs, some of which are more complicated in structure than their eikonal equivalents.
However, they still share the property of being two-external-line irreducible.

In the case of a spinor emitting particle, the additional vertex
\begin{equation}
\frac{i}{2\lambda}\int_0^\infty \sigma^{\mu\nu}F_{\mu\nu}
\label{seagull-nonabel}
\end{equation}
appears in the exponent of Eq.~(\ref{fferm-nonabel}) i.e. the non-abelian analogue of Eq.~(\ref{seagull}). This gives rise 
to both a one- and a two-gluon vertex, derived from the position space expressions
\begin{align}
\frac{i}{\lambda}\int_0^\infty \partial_\mu A^A_\nu(t)\,\sigma^{\mu\nu}\,t^A;\label{chromomag1}\\
\frac12\int_0^\infty \sigma^{\mu\nu}A^A_\mu(t)\, A^b_\nu\,[t^A,t^B](t).\label{chromomag2}
\end{align}
Both of these vertices depend on a single time, and thus are handled similarly to those occurring in Eq.~(\ref{fferm-nonabel3}).
Note that the two-gluon vertex Eq.~(\ref{chromomag2}) has no abelian analogue, which can be seen from the fact that the commutator
$[t^A,t^B]$ vanishes for an abelian gauge field. 

Some comments are in order regarding the color factors associated with the NE webs. We note that the derivation
of the formula for modified color factors in the eikonal case (Eq.~({\ref{modcolsol})) is independent of the eikonal
approximation. Thus, it also applies also in the NE case, where the color factors $C(j)$ associated with each subgraph involving
replica number $j$ are the normal color factors one obtains using the NE Feynman rules.

The above discussion shows that, for the simple hard interaction considered above, a subset of corrections 
exponentiates up to NE order i.e. those associated with soft gauge boson emission outside the hard interaction.
As in the abelian case, one still has to worry about corrections due to eikonal emissions from within the hard interaction
(which give rise to the remainder term in Eq.~(\ref{Mstruc})). This is the subject of the next section.
\subsection{Internal emissions of non-abelian gauge fields}
In the abelian case, we identified a subset of NE corrections which exponentiate (i.e. those arising from soft gauge boson
emissions outside of the hard interaction). There were then remainder terms, which did not have an exponential structure
but could be obtained iteratively to all orders in the perturbation expansion. Here we briefly discuss how this structure can
be generalized to the case of a non-abelian gauge field. As in the previous section, we consider the case where
the matrix element for the hard interaction with no internal emissions has a color singlet structure, with two outgoing 
eikonal lines.

We begin with the non-abelian analogue of the abelian gauge transformation law of Eq.~(\ref{eq:38}), which is
\begin{equation}
f(x_i,p_f;A_\mu)\rightarrow f(x_i,p_f,UA_\mu U^{-1}-iU\partial_\mu U^{-1})=U(x_i)f(x_i,p_f;A_\mu)
\label{nonabelgauge}
\end{equation}
for an external particle, and
\begin{equation}
f(x_i,p_f;A_\mu)\rightarrow f(x_i,p_f,UA_\mu U^{-1}-iU\partial_\mu U^{-1})=f(x_i,p_f;A_\mu)U^{-1}(x_i)
\label{nonabelgauge2}
\end{equation}
for an external antiparticle, where $U(x)=\exp(i\theta^A(x)t^A)$, and the coupling constant $g$ of the 
non-abelian gauge field has been absorbed in $A$.
The hard interaction (in the two eikonal line, color singlet case considered in the previous sections) 
then transforms as
\begin{equation}
H(x_1,x_2,A)\rightarrow H(x_1,x_2,UA_\mu U^{-1}-iU\partial_\mu U^{-1})=U(x_2)H(x_1,x_2)U^{-1}(x_1).
\label{Hnonabeltrans}
\end{equation}
Expanding this to first order in $A$ and $\theta^a(x)$, and using the fact that the latter is arbitrary gives the 
analogue of Eq.~(\ref{eq:40})
\begin{equation}
  \label{eq:40nonabel}
  -\partial_\mu H^\mu (x_1,x_2;x)t^A = i H(x_1,x_2) \sum_j g_j \delta(x-x_j)t^A,
\end{equation}
where $g_j=+1$ for an eikonal particle, and $-1$ for an antiparticle. This has the same interpretation as in the 
abelian case i.e. the amplitude for an internal emission is related to the amplitude with no such emission.
In the non-abelian case, the amplitude for internal emission is proportional to the color matrix $t^A$, as indeed it 
must be from group theory considerations (the quantity $H^\mu$ has one adjoint index and two fundamental indices).

As in the abelian case, Eq.~(\ref{eq:40nonabel}) can be interpreted as an extra vertex for soft gluon emission. This vertex
is located on the Wilson line at $t=0$ i.e. where the eikonal segments $x_1$ and $x_2$ meet. There are terms (also by analogy
with the abelian case of Sec.~\ref{sec:low-theorem}) which correct for the fact that the external lines
do not originate from $x=0$. However, these also take the form of an additional vertex localized at $t=0$. The diagrams
containing these additional vertices do not necessarily exponentiate (as before), and form a remainder term analogously 
to that of Eq.~(\ref{Mstruc}). 

Thus, the all-order structure of matrix elements up to NE order is conceptually equivalent to the abelian case. One has a 
subset of NE corrections which exponentiate (i.e. NE webs), and a set of corrections which form a remainder term which
mixes with the exponentiated NE corrections. The remainder term does not necessarily have a simple structure, but at least has an iterative structure
due to the fact that diagrams with no internal emission are sufficient to generate higher order diagrams 
involving an internal emission.
\section{Discussion}
\label{sec:conclusions}
In this paper, we have considered the issue of soft gauge boson corrections for matrix elements in abelian and non-abelian gauge theories 
from a path integral point of view. This involves considering factorized diagrams involving hard interactions with a given number
of external lines, each of which emits further soft radiation. The propagator for each external line can then be cast into
a first quantized path integral representation, where the integral is a sum over paths for the emitting particle. This path integral
can be performed by expanding about the classical straight line path for the emitter, which corresponds to systematic corrections
to the eikonal approximation. The scattering amplitude then factorizes into a hard interaction plus factors for each external line
which act as source terms for the soft gauge field $A^\mu_s$, where the sources are placed along the external lines. 
In the abelian case, exponentiation of eikonal corrections then follows from the usual exponentiation of disconnected diagrams in 
quantum field theory. Furthermore, a subset of next-to-eikonal corrections can also be shown to exponentiate, and a set of effective 
Feynman rules for radiation in the NE limit is obtained.

The case of a non-abelian gauge field is more complicated, but can be analyzed using the replica trick, in which one considers
an ensemble of $N$ gauge fields. Diagrams which have a term linear in $N$ then exponentiate, and crucially only a subset of
diagrams in the theory have such a property. We considered the simple case of 
two external lines, connected by a hard interaction with color singlet structure. Then the diagrams which exponentiate 
contain sources arising from the replica ordered perturbation theory arising from Eq.~(\ref{prod1}). Those which contribute
at ${\cal O}(N)$ have the property of being two-external line irreducible, and also have (in general) color factors which
differ from those of the corresponding diagrams in the original theory. These diagrams are then precisely the webs of
\cite{Frenkel:1984pz,Berger:2003zh}. As in the abelian case, a subset of NE corrections also exponentiates, and the exponent
contains a sum of webs up to NE level.

In both the abelian and non-abelian cases, there are NE corrections which do not exponentiate, and which form a remainder term
such that the total matrix element (up to NE order) has the form shown in Eq.~(\ref{Mstruc}). These terms are associated with 
Low's theorem, and the relevant diagrams involve contractions between eikonal photons or gluons on the external lines, and
a NE vertex localized at the cusp at which the outgoing eikonal lines meet. Furthermore, these terms have an iterative structure
in perturbation theory, in that the extra vertices that contribute can be related to diagrams at a lower order in the 
perturbation expansion.
 
A comment is in order regarding the nature of NE exponentiation. Up to NE order, Eq.~(\ref{Mstruc}) can be expanded to give
\begin{equation}
{\cal M}=\exp\left[{\cal M}^{\text{E}}\right]\left(1 + {\cal M}_r+{\cal M}^{\text{NE}}\right)
\label{Mstruc2}
\end{equation}
i.e. one may either consider the NE terms arising from eikonally factorized diagrams as being in the exponent, or
kept to linear order. This masks the fact that the terms in ${\cal M}^{\text{NE}}$ genuinely do 
exponentiate, whereas the remainder terms in ${\cal M}_r$ do not. However, it is true that the exponentiated NE terms
lead to NNE, NNNE etc. contributions which would then mix with higher order (in the eikonal expansion) remainder terms.
The exponentiated form of Eq.~(\ref{Mstruc2}) is particularly useful if the contribution from ${\cal M}^{\text{NE}}$ (when
exponentiated) gives the dominant contribution to higher order terms in the eikonal expansion. Whether or not this is the
case is presumably process dependent. 

The proof of exponentiation in the abelian case, as presented here, clearly generalizes to higher numbers of eikonal lines.
However, in the non-abelian case we considered only the simplest possible hard interaction, namely that with two external lines
with a color singlet structure. This could be easily related to a single Wilson line with a cusp. The general case of higher numbers
of external lines is more complicated due to the color structures involved. 
Nevertheless, the methods introduced in this paper may provide a useful starting point in addressing NE corrections
in these situations.

We have only considered matrix elements in this paper. Thus, any exponentiation of NE corrections pertains only before any integration
over the phase space of the final state gauge bosons has been performed. In the strict eikonal approximation, the phase space
factorizes into a product of single particle phase spaces (i.e. conservation of energy is a subleading effect), thus exponentiation
in the matrix element implies exponentiation of soft logarithms in differential (but partially integrated) cross-sections. This is not
necessarily the case beyond the eikonal approximation, where one expects NE corrections resulting from the eikonal matrix element 
with integration over the full phase space. Na\"{i}vely, one expects a given differential cross-section (e.g. in some variable $\xi$
related to the total energy fraction carried by soft gluons) to have the form
\begin{equation}
\frac{d\sigma}{d\xi}=\int d\text{PS}^{(\text{E})}\,|{\cal M}^{(\text{E})}|^2+\left[\int d\text{PS}^{(\text{E})}\,|{\cal M}^{(\text{NE})}|^2
+\int d\text{PS}^{(\text{NE})}\,|{\cal M}^{(\text{E})}|^2\right]+{\cal O}(\text{NNE}).
\label{phsp}
\end{equation}
Here ${\cal M}^{(\text{E, NE})}$ denote the eikonal and next-to-eikonal matrix elements respectively, and $d\text{PS}^{(\text{E})}$ the 
eikonal phase space, consisting of a factorized product of one-particle phase spaces. The first term in Eq.(\ref{phsp}) is then of eikonal
order, and the bracketed term is NE, where $d\text{PS}^{(\text{NE})}$ represents that part of the multi-gluon phase space which 
implements next-to-eikonal corrections (i.e. subleading terms in $\xi$). 
The precise nature of this latter term is unclear, and an investigation of its effect is deferred
to a future publication. Nevertheless, all of the ingredients for the first bracketed term in Eq.~(\ref{phsp}) are contained in this paper.

To conclude, the path integral methods used in this paper provide a new viewpoint for the exponentiation of soft radiative corrections to matrix elements,
in both abelian and non-abelian gauge theories. In particular, the discussion of webs is rephrased such that a closed form solution for the modified
color factors can be given. Furthermore, the approach naturally encompasses the exponentiation of classes of next-to-eikonal corrections.
This approach should prove fruitful in the further investigation of soft radiative corrections to all orders in perturbation theory.
\acknowledgments 
We would like to thank Jan-Willem van Holten and Lorenzo Magnea for
valuable discussions. The work of EL and CDW is supported by the 
Netherlands Foundation for Fundamental Research of Matter (FOM) 
and the National Organization for Scientific Research (NWO).

\appendix
\section{Exponentiation of disconnected diagrams}
\label{expproof}
Here we briefly prove the exponentiation of disconnected diagrams in quantum field theory,
using the {\it replica trick} of statistical physics. Although we consider a single self-interacting
scalar field $\phi$, the proof generalizes easily to other systems. 

The Green's functions of a given quantum field theory are described 
by the generating functional
\begin{equation}
Z[J]=\int\measure{\phi}e^{iS[\phi]+i\int J\phi},
\label{Z}
\end{equation}
where $J$ is a source for the field $\phi$, and
$S$ is the classical action. Now consider defining $N$ replicas of the theory,
involving fields $\phi_i$ ($i\in\{1,\ldots,n\}$). This has generating functional
\begin{equation}
Z_N[J]=\int\measure{\phi_1}\ldots\measure{\phi_N}e^{iS[\phi_1]+i\int J\phi_1}\ldots e^{iS[\phi_N]+i\int J\phi_N},
\label{ZN}
\end{equation}
which clearly satisfies
\begin{equation}
Z_N[J]=(Z[J])^N.
\label{ZN2}
\end{equation}
The Feynman rules for each field are similar, and there are no interactions between the fields. 
Thus, there can be no more than one field in each connected Feynman diagram, and connected diagrams
therefore have $N$ copies. By the same reasoning, disconnected diagrams containing $n\geq2$ 
constituent parts have $N^n$ copies. It follows that
\begin{equation}
\sum G_c\propto N,
\label{Gcsum}
\end{equation}
where $G_c$ denotes a connected diagram. Furthermore, no disconnected diagrams contribute 
terms proportional to $N$. From Eq.~(\ref{ZN2}) one has
\begin{equation}
Z_N[J]=1+N\log(Z[J])+{\cal O}(N^2),
\label{ZN3}
\end{equation}
and comparing Eqs.~(\ref{Gcsum}, \ref{ZN3}) gives
\begin{equation}
\sum G_c=\log(Z[J]).
\label{Gcsum2}
\end{equation}
Finally, one writes this as:
\begin{equation}
Z[J]=\exp\left[\sum G_c\right]
\label{Z2}
\end{equation}
and sets $N=1$. This is the statement that disconnected diagrams exponentiate, as required.

The above proof shortcuts the nontrivial combinatoric fact that the symmetry factor associated
with a disconnected Feynman graph is the product of the symmetry factors of each of the 
constituent connected graphs. The results of this paper demonstrate that the same combinatoric
reasoning applies to the exponentiation of soft radiative corrections from fast-moving external
particles.

\section{Next-to-eikonal Feynman rules}
\label{sec:next-eikonal-feynman}
In this appendix, we show how to derive effective Feynman rules for the eikonal and next-to-eikonal
approximations in the path integral approach discussed in section \ref{sec:next-eikon-expon}, and
show explicitly that these agree with the rules one obtains by starting from exact perturbation theory 
before taking the NE limit.

We begin with the conventional perturbative approach, and consider the case of a charged scalar
interacting with a background gauge field. The theory and its Green functions are defined via the 
generating functional
\begin{equation}
\label{eq:29}
Z[J,J^*]= \int\measure{\phi}\measure{\phi^*}\,
\exp\Big[i\int d^dx\left[-(D_\mu\phi)^*D^\mu\phi-m^2\phi^*\phi +J \phi^* + J^*\phi \right]\Big]\,.
\end{equation}
where $D_\mu = \partial_\mu-iA_\mu$. From this functional one derives
the Feynman rules
\begin{equation}
  \begin{split}
    \includegraphics[scale=.6]{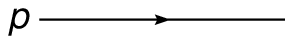} & \qquad \frac1{i (p^2+m^2-i\varepsilon)},\\
    \includegraphics[scale=.6]{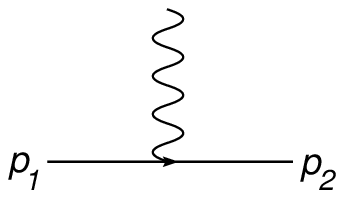}& \qquad i (p_1+p_2)^\mu,\\
    \includegraphics[scale=.6]{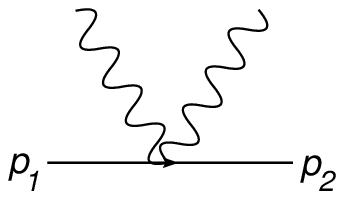}& \qquad -i \,2\eta^{\mu\nu}.
  \end{split}
\end{equation}
Note the symmetry factor in the seagull vertex. 
With these rules one can describe emissions from a charged scalar line,
and from them we can derive Feynman rules for the eikonal and NE approximation. To 
do so, we shall take $m=0$ and use  $\lambda$-scaling as discussed in
section \ref{sec:soft-emiss-scatt}. That is, we consider an external particle
of 4-momentum $p=\lambda n$, and consider the limit $\lambda\rightarrow\infty$.
As discussed in section \ref{sec:soft-emiss-scatt}, this corresponds to the eikonal approximation, 
with the first subleading corrections in $\lambda$ representing the NE approximation. 

We first consider the propagator-vertex combination for one-photon emission. 
The diagram and corresponding expression is 
\begin{equation}
  \label{eq:23}
    \includegraphics[scale=0.4]{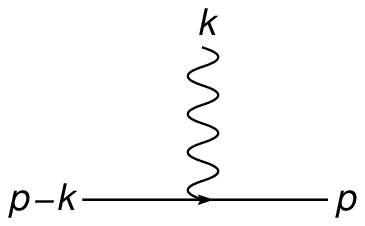}\qquad \qquad \frac1{i (p-k)^2}i(2 p - k)^\mu\,,
\end{equation}
where we have taken this diagram to represent the combined vertex and propagator factors (to the left of the vertex).
Setting $p=\lambda n$ and expanding Eq.~(\ref{eq:23}) to  $\mathcal{O}(1/\lambda)$  yields
\begin{equation}
\label{eq:30}
 -\frac{n^\mu}{n\cdot k}+\frac1\lambda
\left(\frac{k^\mu}{2 n\cdot k}-k^2\frac{n^\mu}{2(n\cdot k)^2}\right).
\end{equation}
We recognize the first term as the eikonal approximation, and the
remainder as the NE contribution. Each term in Eq.~\eqref{eq:30} is to be contracted 
with a background gauge field, so that we can treat each
as a one-photon source.  We represent these sources graphically as
\begin{equation}\label{eq_ne_1}
  \begin{split}
    \includegraphics[scale=.6]{eikonal-1vertex.eps} &\qquad \text{(1)} \qquad -\frac{p^\mu}{p \cdot k},\\
    \includegraphics[scale=.6]{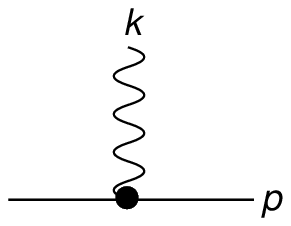}&\qquad\text{(2a)} \qquad \frac{k^\mu}{2 p\cdot k},\\
    \includegraphics[scale=.6]{NE-1vertex.eps}&\qquad\text{(2b)} \qquad -k^2\frac{p^\mu}{2(p\cdot k)^2}.
  \end{split}
\end{equation}
Note that here and in following graphs, we replace $n\rightarrow p/\lambda$ so that the Feynman rules
are given in terms of the physical momenta.
We now consider possible two-photon sources. Starting with the seagull term, 
we take the following propagator-vertex combination:
\begin{equation}
  \includegraphics[scale=.6]{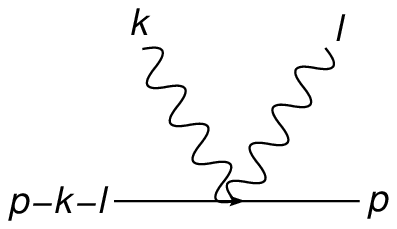}\qquad \frac1{i (p-k-l)^2}(-i2\eta^{\mu\nu}).
\end{equation}
Scaling $p$ and gathering terms up to $\mathcal{O}(1/\lambda)$ yields
\begin{equation}
\label{eq:27}
  \frac1\lambda\frac{\eta^{\mu\nu}}{n\cdot (k+l)}.
\end{equation}
Evidently, at the eikonal level the seagull term is absent. One expects this given that 
there is no such seagull vertex in the exact Feynman rules for a fermionic emitting particle,
and, as is well known, in the eikonal approximation the emitted radiation is insensitive 
to the particle's spin.

We next examine the contribution of two individual
photon emissions, as shown in Fig. \ref{Fig:A13}. At eikonal level these diagrams give
a contribution
\begin{equation}
\left(-\frac{n^\nu}{n\cdot (k+l)}\right)\left(-\frac{n^\mu}{n\cdot k}\right)
+\left(-\frac{n^\nu}{n\cdot(k+l)}\right)\left(-\frac{n^\mu}{n\cdot(l)}\right),
\label{twoeik}
\end{equation}
which one may rearrange to give:
\begin{equation}
\left(-\frac{n^\nu}{p\cdot k}\right)\left(-\frac{n^\mu}{p\cdot l}\right).
\label{tweik2}
\end{equation}
The contribution from the soft emissions explicitly factorizes into a product of uncorrelated 
emissions, as is well-known.

At ${\cal O}(1/\lambda)$, corresponding to the NE limit, one must sum over all possible insertions
of a NE one-photon emission vertex in Fig. (\ref{Fig:A13}). There are possible vertices, as given in
Eq.~(\ref{eq_ne_1}). The vertex (2a) yields an expression
\begin{figure}
  \begin{center}
   \includegraphics[height=3cm]{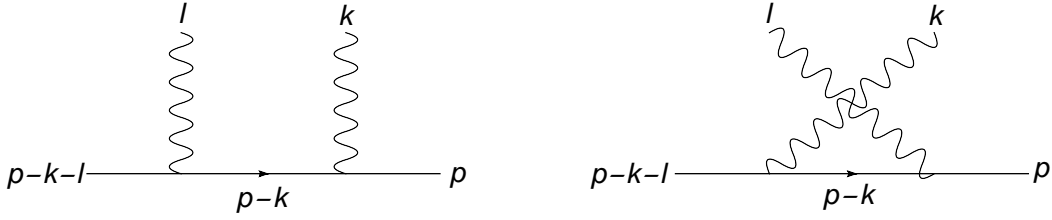}
\caption{Diagrams contributing to two photon emission. At NE level, one must sum over all possible
insertions of a NE one-photon emission vertex, as given in Eq.~(\ref{eq_ne_1}).}
\label{Fig:A13}
  \end{center}
\end{figure}
\begin{multline}
\label{eq:25}
  \left(-\frac{n^\nu}{n\cdot (k+l)}\right)\left(\frac{k^\mu}{2 n\cdot
      k}\right)+
  \left(\frac{(2k+l)^\nu}{2n\cdot (k+l)}\right)\left(-\frac{n^\mu}{n\cdot k}\right)+\\
  \left(-\frac{n^\mu}{n\cdot (k+l)}\right)\left(\frac{l^\nu}{2 n\cdot
      l}\right)+ \left(\frac{(k+2l)^\mu}{2n\cdot
      (k+l)}\right)\left(-\frac{n^\nu}{n\cdot l}\right)\,.
\end{multline}
which can be rearranged to give
\begin{equation}
\label{eq:26}
  \left(-\frac{n^\nu}{n\cdot l}\right)\left(\frac{k^\mu}{2 n\cdot k}\right)+
  \left(-\frac{n^\mu}{n\cdot k}\right)\left(\frac{l^\nu}{2 n\cdot l}\right)-
  \frac{l^\mu n^\nu n\cdot k+k^\nu n^\mu n\cdot l}{n\cdot (k+l)
 n\cdot k n\cdot l}.
\end{equation}
Notice that the first two terms correspond to two uncorrelated 
NE emissions, while the last term represents a correlated two-photon 
emission (i.e. is non-factorizable into terms dependent on a single 
photon momentum). The NE vertex (2b) gives a contribution
\begin{multline}
\label{eq:24}
  \left(-\frac{n^\nu}{n\cdot (k+l)}\right)\left(-k^2 \frac{n^\mu}{2
      (n\cdot k)^2}\right)+
  \left(-(k+l)^2 \frac{n^\nu}{2(n\cdot (k+l))^2}\right)\left(-\frac{n^\mu}{n\cdot k}\right)+\\
  \left(-\frac{n^\mu}{n\cdot (k+l)}\right)\left(-l^2 \frac{n^\nu}{2
      (n\cdot l)^2}\right)+ \left(-(k+l)^2\frac{n^\mu}{2 (n\cdot
      (k+l))^2}\right)\left(-\frac{n^\nu}{n\cdot l}\right)\,,
\end{multline}
which can be rewritten as
\begin{equation}
\label{eq:28}
  \left(-\frac{n^\nu}{n\cdot l}\right)\left(-k^2\frac{n^\mu}{2 (n\cdot k)^2}\right)+
  \left(-\frac{n^\mu}{n\cdot k}\right)\left(-l^2\frac{n^\nu}{2 (n\cdot l)^2}\right)+
  \frac{n^\mu n^\nu k\cdot l}{n\cdot (k+l) n\cdot k n\cdot l}\,.
\end{equation}
Again this is the sum of an uncorrelated part, and a term implementing correlated photon emission.
The various correlated contributions given in Eqs.~(\ref{eq:27}), (\ref{eq:26}) and (\ref{eq:28}) 
can be represented as new two-photon vertices, and are given respectively by
\begin{equation}\label{eq_ne_2}
  \begin{split}
    \includegraphics[scale=.6]{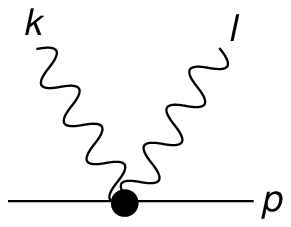} &\qquad\text{(3)} \qquad +\frac{\eta^{\mu\nu}}{p\cdot (k+l)},\\
    \includegraphics[scale=.6]{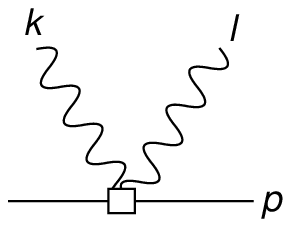} &\qquad\text{(4a)} \qquad -\frac{l^\mu p^\nu p\cdot k+k^\nu p^\mu p\cdot l}{p\cdot (k+l) p\cdot k p\cdot l},\\
    \includegraphics[scale=.6]{NE-2vertex2.eps} &\qquad\text{(4b)} \qquad +\frac{p^\mu p^\nu k\cdot l}{p\cdot (k+l)
      p\cdot k p\cdot l}.
  \end{split}
\end{equation}
We have shown that these vertices apply when two photons are emitted next to the on-shell eikonal
line. From our analysis in this appendix, it does not necessarily follow that these vertices 
apply to emissions anywhere on the external line,
at all orders of perturbation theory. That this is indeed the case is clear when one rederives these
vertices using the path integral methods described in this paper, and we defer a full proof within 
conventional perturbation theory to a forthcoming paper~\cite{LMSW}. Here we will only demonstrate 
that these terms are precisely reproduced 
in our path-integral formalism.  The path integral representation of a charged 
scalar coupled to a background gauge field is given by Eq.~(\ref{eq:22}) as 
\begin{multline}
\label{eq:31}
f(\infty)=  \int_{x(0)=0} \measure{x}\, \exp\Big[i\int_0^\infty dt \left(\frac\lambda2\dot{x}^2+(n+\dot{x})\cdot A(x_i+nt+x)\right.\\
     \left. +\frac{i}{2\lambda}\partial\cdot A(x_i+p_f t+x)\right)\Big]\,.
\end{multline}
Our task is to derive the photon source terms from this expression. To this end we need
to determine the propagator and vertices for the $x$ field, and their scaling with $\lambda$.

The $x$ kinetic term is given by
\begin{equation}
\label{eq:32}
  -\int_0^\infty dt \frac12 x(t)\,\left(i\lambda\frac{\partial^2}{\partial t^2}\right)\,x(t)\,,
\end{equation}
The propagator for $x$ is given by the inverse of the quadratic operator in Eq.~(\ref{eq:32}),
which is found to be
\begin{equation}
\label{eq:33}
  G(t,t')=\frac i\lambda \min(t,t')\,.
\end{equation}
Note that it is symmetric, proportional to $1/\lambda$ (thus is of NE order), and satisfies the condition  $G(0,t')=0$. 
Other two-point correlators of $x$ and $\dot{x}$, which we need below, are
\begin{equation}
\label{correlators}
  \begin{split}
    \langle x(t) x(t')\rangle=G(t,t')=\frac i\lambda \min(t,t'),\\
    \langle \dot{x}(t) x(t')\rangle=\frac{\partial G(t,t')}{\partial t}=\frac i\lambda \theta(t'-t),\\
    \langle \dot{x}(t) \dot{x}(t')\rangle=\frac{\partial^2 G(t,t')}{\partial t\partial t'}=\frac i\lambda \delta(t'-t)\,.
  \end{split}
\end{equation}
We will also need the properties of the equal time correlator $\langle\dot{x(t)}x(t)\rangle$. Using the discretization
of space-time adopted throughout this paper, the derivative $\dot{x}(t)$ is given by
\begin{equation}
\lim_{\epsilon\downarrow 0}\epsilon\frac{x(t+\epsilon)-x(t)}{\epsilon}
\label{xdot}
\end{equation}
and thus one has
\begin{equation}
\langle\dot{x(t)}x(t)\rangle=\frac{i}{\lambda}\lim_{\epsilon\downarrow 0}\frac{\text{min}(t+\epsilon,t)-\text{min}(t,t)}{\epsilon}=0.
\label{xdot2}
\end{equation}
The vertices involving the $x$ field can be obtained by Taylor expansion of
the other terms in Eq. (\ref{eq:31}). Due to the subleading nature of the $x$ propagator in the eikonal limit, 
we shall need them to second order in $x$ or $\dot{x}$ only for a NE analysis. 
The terms without a power of $x$ are
\begin{equation}
\label{eq:34}
  \begin{split}
    i\int_0^\infty dt n\cdot A(nt)&=-\int \frac{d^dk}{(2\pi)^d} \frac{n^\mu}{n\cdot k} \tilde{A}_\mu(k),\\
    -\frac1{2\lambda}\int_0^\infty dt \partial\cdot
    A(nt)&=\frac1{2\lambda} \int \frac{d^dk}{(2\pi)^d}
    \frac{k^\mu}{n\cdot k} \tilde{A}_\mu(k),
  \end{split}
\end{equation}
where we have also represented the terms in momentum space. Note that these
$A$ source terms correspond to the vertex in Eq.~(\ref{eq_ne_1}) (2a). 
Terms with one power of $x$ are
\begin{equation}\label{eq:one}
  \begin{split}
    i\int_0^\infty dt \dot{x}^\mu A_\mu(nt)&=
    \int \frac{d^dk}{(2\pi)^d} \tilde{A}_\mu(k)\int_0^\infty dt\,i\dot{x}^\mu(t) e^{i (n\cdot k)t}\quad \clubsuit\\
    i\int_0^\infty dt n^\mu \partial_\nu A_\mu(nt)x^\nu(t)&=-\int
    \frac{d^dk}{(2\pi)^d} n^\mu \tilde{A}_\mu(k) k_\nu \int_0^\infty dt\,
    x^\nu(t) e^{i (n\cdot k)t}\quad \diamondsuit
  \end{split}
\end{equation}
To distinguish the vertices in our discussion below, we have labeled them with symbols.
Terms with two powers of $x$ are
\begin{equation}
\label{eq:35}
  \begin{split}
    i\int_0^\infty dt \dot{x}^\mu \partial_\nu A_\mu(nt) x^\nu&=
    -\int \frac{d^dk}{(2\pi)^d} \tilde{A}_\mu(k) k_\nu \int_0^\infty dt \dot{x}^\mu(t) x^\nu(t) e^{i (n\cdot k)t} \quad  \heartsuit\\
    \frac i2 \int_0^\infty dt n^\mu \partial_\nu \partial_\kappa
    A_\mu(nt)x^\nu(t)x^\kappa(t)&= -\frac i2 \int \frac{d^dk}{(2\pi)^d} n^\mu
    \tilde{A}_\mu(k) k_\nu k_\kappa \int_0^\infty dt x^\nu(t)
    x^\kappa(t) e^{i (n\cdot k)t}\,.\quad \spadesuit
  \end{split}
\end{equation}
The term $\spadesuit$ is quadratic in $x$, and thus the factor of $1/2$ does not appear in the resulting vertex.

The next step is to carry out the $x$ path-integral. This amounts to using the Feynman 
rules in Eqs.~(\ref{eq:33} - \ref{eq:35}), keeping terms to $\mathcal{O}(1/\lambda)$ 
\begin{multline}
\label{eq:36}
  \int_{x(0)=0} \measure{x} \,\exp\Big[i\int_0^\infty dt
    \left(\frac\lambda2\dot{x}^2+(n+\dot{x})\cdot A(x_i+nt+x)
      +\frac{i}{2\lambda}\partial\cdot A(x_i+p_f t+x)\right)\Big]=\\
  \exp\left[-\int \frac{d^dk}{(2\pi)^d} \frac{n^\mu}{n\cdot k}
    \tilde{A}_\mu(k)+ \frac1{2\lambda} \int \frac{d^dk}{(2\pi)^d}
    \frac{k^\mu}{n\cdot k} \tilde{A}_\mu(k)+\sum
    \includegraphics[scale=0.3]{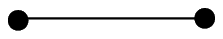}+\sum\includegraphics[scale=0.3]{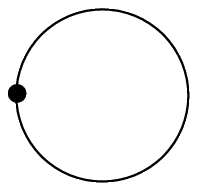}\right].
\end{multline}
We have written the result as the exponent of connected diagrams, where terms beyond ${\cal O}(1/\lambda)$
are neglected. The first two terms (\ref{eq:34})  are the vertices (1) and (2a) of \eqref{eq_ne_1}.  
The third and fourth terms represent tree and loop graphs, with a sum over all possible insertions of the 
vertices denoted above by $\clubsuit$, $\diamondsuit$, $\heartsuit$ and $\spadesuit$. In the tree graph,
there are three different combinations of one-$x$ vertices
from \eqref{eq:one}.  The $\clubsuit-\clubsuit $ combination together with a two-$x$ correlator gives
\begin{multline}
  \frac12 \int \frac{d^dk}{(2\pi)^d} \frac{d^dk}{(2\pi)^d}
  \tilde{A}_\mu(k) \tilde{A}_\nu(l)
  \int_0^\infty dt\, dt'\, \frac{-i}\lambda \delta(t-t') \eta^{\mu\nu} e^{i(n\cdot k t+n\cdot l t')}=\\
  \frac12 \int \frac{d^dk}{(2\pi)^d} \frac{d^dk}{(2\pi)^d}
  \tilde{A}_\mu(k) \tilde{A}_\nu(l) \frac{\eta^{\mu\nu}}{\lambda
    n\cdot(k+l)}.
\end{multline}
This precisely reproduces source term (3) of Eq.~\eqref{eq_ne_2}. 
The $\clubsuit-\diamondsuit$ combination gives
\begin{multline}
  \int \frac{d^dk}{(2\pi)^d} \frac{d^dk}{(2\pi)^d} \tilde{A}_\mu(k)
  \tilde{A}_\nu(l) n^\nu(-k_\kappa)
  \int_0^\infty dt\, dt'\, i \left(\frac i\lambda \delta(t-t')\right) \eta^{\mu\kappa} \frac{e^{i(n\cdot k t+n\cdot l t')}}{-i n\cdot l}=\\
  \int \frac{d^dk}{(2\pi)^d} \frac{d^dk}{(2\pi)^d} \tilde{A}_\mu(k)
  \tilde{A}_\mu(l) \frac{-n^\nu l^\mu}{\lambda n\cdot l n\cdot(k+l)}\,,
\end{multline}
which can be rewritten as
\begin{equation}
  \frac12 \int \frac{d^dk}{(2\pi)^d} \frac{d^dk}{(2\pi)^d} \tilde{A}_\mu(k) \tilde{A}_\nu(l)
  \left(-\frac{n^\nu l^\mu n\cdot k+n^\mu k^\nu n\cdot l}{\lambda n\cdot l n\cdot k n\cdot(k+l)}\right).
\end{equation}
This is source term (4a) of \eqref{eq_ne_2}. The $\diamondsuit-\diamondsuit$ 
combination gives
\begin{multline}
  \frac12 \int \frac{d^dk}{(2\pi)^d} \frac{d^dk}{(2\pi)^d}
  \tilde{A}_\mu(k) \tilde{A}_\nu(l) n^\mu\, n^\nu\, k_\rho\, l_\sigma
  \int_0^\infty dt\, dt'\, \left(\frac i\lambda \delta(t-t')\right)
  \eta^{\rho\sigma}
  \frac{e^{i(n\cdot k t+n\cdot l t')}}{(-i n\cdot k)(-i n\cdot l)}=\\
  \frac12 \int \frac{d^dk}{(2\pi)^d} \frac{d^dk}{(2\pi)^d}
  \tilde{A}_\mu(k) \tilde{A}_\mu(l) n^\mu n^\nu \frac{k\cdot
    l}{\lambda n\cdot l n\cdot k n\cdot(k+l)}\,.
\end{multline}
which produces term (4b) of \eqref{eq_ne_2}.

The loop graph in Eq.~(\ref{eq:36}) has in principle two possible choices of 
vertex from \eqref{eq_ne_2}. However, the $\heartsuit$
vertex does not actually contribute, as it involves the equal time correlator
$\langle\dot{x}(t)x(t)\rangle$, which was shown above to be zero. 
The $\spadesuit$ vertex, however,  does contribute and gives
\begin{equation}
  -\frac12 i \int \frac{d^dk}{(2\pi)^d} \frac{d^dk}{(2\pi)^d} \tilde{A}_\mu(k) k_\rho k_\sigma
  \int dt\,\frac i\lambda\, t\, e^{i n\cdot k t}=\\
  -\frac12 \frac{d^dk}{(2\pi)^d} \tilde{A}_\mu(k) \frac{k^2}{(n\cdot k)^2}.
\end{equation}
This finally yields source term (2b) of \eqref{eq_ne_1}. 

We conclude that, to next-to-eikonal order, the one- and two-photon source terms found
by approximations in standard perturbation theory are precisely
reproduced in our first-quantized path-integral approach. The considerations of section
\ref{sec:soft-emiss-scatt} then show that NE corrections from eikonally factorized diagrams
exponentiate to all orders.

We have considered the case of scalar emitting particles in the above discussion. However, 
things proceed similarly for the spinor case, with the only modification arising due to the presence
of an additional term in the exponent of Eq.~(\ref{eq:31})
\begin{equation}
\frac{i}{2\lambda}\int_0^\infty \sigma^{\mu\nu}F_{\mu\nu}=-\frac{1}{\lambda}\int \frac{d^dk}{(2\pi)^d} k_\nu\,[\gamma^\nu,\gamma^\mu]\,\tilde{A}_\mu(k),
\label{seagull}
\end{equation}
where the right-hand-side corresponds to the momentum space vertex
\begin{equation}
    \includegraphics[scale=.6]{NE-1vertex.eps}\qquad -k_\nu\,[\gamma^\nu,\gamma^\mu].\\
\label{spin1v}
\end{equation}
This is ${\cal O}(1/\lambda)$, as expected from the fact
that a magnetic moment vertex only contributes for particles having non-zero spin, and radiation in the strictly eikonal limit is insensitive
to the spin of the emitting particle.

The above discussion assumes an abelian gauge field. The non-abelian generalization is discussed in Sec.~\ref{sec:non-abelian-gauge}.
\section{Matrix elements with internal emissions}
\label{sec:low-example}
In section \ref{sec:low-theorem}, we show from our path integral representation of the scattering amplitude
how NE corrections arising from soft photon emissions from within the hard interaction can be related to the 
hard interaction with no emissions, which is part of the content of the Low-Burnett-Kroll 
theorem~\cite{Low:1958sn,Burnett:1967km} (see also \cite{DelDuca:1990gz}).
To clarify this discussion, we here present how one would obtain a 
similar result using traditional Feynman diagram methods, in the case of two scalar lines. 
Our presentation follows that of e.g.~\cite{Sterman:1994ce}. 

We consider the momentum-space amplitude $\Gamma^\mu$ shown pictorially in Fig.~\ref{Gam}, and corresponding 
to a single gluon emission emitted from a graph containing a given hard interaction with two
external scalar particles. 
\begin{figure}
\begin{center}
\scalebox{0.8}{\includegraphics{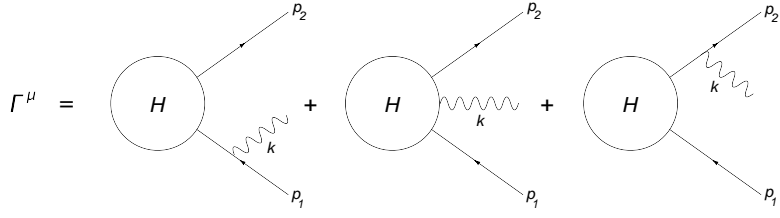}}
\caption{Amplitude $\Gamma^\mu$ corresponding to all possible soft emissions from a graph containing a given hard interaction,
in the simple case of two outgoing scalar particles. The Lorentz index corresponds to the emitted photon.}
\label{Gam}
\end{center}
\end{figure}
We first define $\Gamma\equiv\Gamma(p_1^2,p_2^2,p_1\cdot p_2)$ to be the hard interaction amplitude with no photon emission.
Then using the normal Feynman rules for scalar electrodynamics, one may write $\Gamma^\mu$ as 
\begin{equation}
\Gamma^\mu=\frac{(2p_1-k)^\mu}{-2p_1\cdot k}\Gamma[(p_1-k)^2,p_2^2,(p_1-k)\cdot p_2]+\frac{(2p_2+k)^\mu}{2p_2\cdot k}\Gamma[p_1^2,(p_2+k)^2,p_1\cdot(p_2+k)]+\Gamma^\mu_{int},
\label{Gam1}
\end{equation}
where we have assumed light-like external particles, and $\Gamma^\mu_{int}$ is the amplitude for emission 
from within the hard interaction, as represented by the second diagram on the right-hand-side in figure \ref{Gam}.
Also, we have neglected coupling constants for brevity.
In the limit where $k$ is soft, one may Taylor expand Eq.~(\ref{Gam1}) to next-to-leading order in $k$ (corresponding
to the NE approximation) to obtain
\begin{align}
\Gamma^\mu&=\frac{2p_1^\mu}{-2p_1\cdot k}\left[\Gamma-2p_1\cdot k\frac{\partial\Gamma}{\partial p_1^2}
-p_2\cdot k\frac{\partial\Gamma}{\partial p_1\cdot p_2}\right]+k^\mu\left(\frac{1}{2p_1\cdot k}+\frac{1}{2p_2\cdot k}\right)\notag\\
&+\frac{2p_2^\mu}{2p_2\cdot k}\left[\Gamma+2p_2\cdot k\frac{\partial\Gamma}{\partial p_2^2}+p_1\cdot k\frac{\partial\Gamma}{\partial p_1\cdot p_2}\right]+\Gamma^\mu_{int}.
\label{Gam2}
\end{align}
The gauge invariance condition $k_\mu\Gamma^\mu=0$ implies
\begin{equation}
k_\mu\Gamma^\mu_{int}=-2p_2\cdot k\frac{\partial\Gamma}{\partial p_2^2}-2p_1\cdot k\frac{\partial\Gamma}
{\partial p_1^2}-k\cdot(p_1+p_2)\frac{\partial\Gamma}{\partial p_1\cdot p_2}.
\label{Gam3}
\end{equation}
This must be true for arbitrary $k$, so that one may remove factors of $k^\mu$ in Eq.~(\ref{Gam3}). Then one may substitute the resulting
form of $\Gamma^\mu$ back into Eq.~(\ref{Gam}) to obtain
\begin{equation}
\Gamma^\mu=\left[\frac{(2p_1-k)^\mu}{-2p_1\cdot k}+\frac{(2p_2+k)^\mu}{2p_2\cdot k}\right]\Gamma
+\left[\frac{p_1^\mu(k\cdot p_2-k\cdot p_1)}{p_1\cdot k}+\frac{p_2^\mu(k\cdot p_1-k\cdot p_2)}{p_2\cdot k}
\right]\frac{\partial\Gamma}{\partial p_1\cdot p_2}.
\label{Gam4}
\end{equation}
Identifying $p_i=n_i$ in the notation of section \ref{sec:low-theorem}, one sees that the first term of Eq.~(\ref{Gam4})
is the contribution one expects from the effective Feynman rules for soft emission up to NE order. Also, the second term
in Eq.~(\ref{Gam4}) is precisely the contribution contained in Eq.~(\ref{eq:47b}). Furthermore, when $p_1^2=p_2^2=0$, one may 
simplify Eq.~(\ref{Gam3}) so that one obtains
\begin{equation}
\Gamma^\mu_{int}=-p_1^\mu\frac{\partial\Gamma}{\partial p_1\cdot p_2}-p_2^\mu\frac{\partial\Gamma}{\partial p_1\cdot p_2}.
\label{Gam5}
\end{equation}
Rewriting
\begin{equation}
p_i^\mu\frac{\partial}{\partial p_i\cdot p_j}=\frac{\partial}{\partial p_i^\mu},
\label{derivs}
\end{equation}
one finds
\begin{equation}
\Gamma^\mu_{int}=-\sum_i \frac{\partial\Gamma}{\partial p_i^\mu},
\label{Gam6}
\end{equation}
which is indeed a special case of Eq.~(\ref{eq:43}).

\bibliographystyle{JHEP}
\bibliography{refs.bib}

\end{document}